\documentclass[prd,preprint,superscriptaddress,showpacs,nofootinbib,tightenlines,amsmath]{revtex4-1}
\usepackage{graphicx}
\usepackage{epsfig}
\usepackage{slashed}
\usepackage{braket}
\usepackage{nicefrac}
\usepackage{bbm}
\usepackage{amsmath,amssymb}
\usepackage{color}
\usepackage{MnSymbol}

\def\no{\nonumber}

 \def\la{\langle}
 \def\ra{\rangle}

 \def\chip{\chi_+}
 
 \def\chim{\chi_-}

\newcommand{\bda}{\begin{\displaymath}\begin{array}{rl}}
\newcommand{\eda}{\end{array}\end{displaymath}}
\newcommand{\be}{\begin{equation}}
\newcommand{\ee}{\end{equation}}
\newcommand{\bdm}{\begin{displaymath}}
\newcommand{\edm}{\end{displaymath}}
\newcommand{\bea}{\begin{eqnarray}}
\newcommand{\eea}{\end{eqnarray}}

\newcommand{\myTR}{\hline\hline}
\newcommand{\myMR}{\hline}
\newcommand{\myBR}{\hline\hline}

\begin{document}
\title{$\eta^{(')}\to\pi^+\pi^-\gamma^{(\ast)}$ in large-$N_c$ chiral perturbation theory}
\author{P.~Bickert}
\affiliation{Institut f\"ur
Kernphysik, Johannes Gutenberg-Universit\"at Mainz, D-55099 Mainz,
Germany}
\affiliation{Fraunhofer-Institut f\"ur Techno- und Wirtschaftsmathematik ITWM, D-67663 Kaiserslautern, Germany}
\author{S.~Scherer}
\affiliation{Institut f\"ur
Kernphysik, Johannes Gutenberg-Universit\"at Mainz, D-55099 Mainz,
Germany}
\date{\today}
\begin{abstract}

We present a calculation of the decays $\eta^{(')}\to\pi^+\pi^-\gamma^{(\ast)}$ at the one-loop level up to and including next-to-next-to-leading order (NNLO) in large-$N_c$ chiral perturbation theory. The numerical evaluation of the results is performed successively at LO, NLO, and NNLO, fitting the relevant low-energy constants to the available experimental data. We discuss the widths and decay spectra of $\eta^{(')}\to\pi^+\pi^-\gamma$ as well as $\eta^{(')}\to\pi^+\pi^-l^+l^-$, with $l=e,\ \mu$.

\end{abstract}
\maketitle

\section{Introduction}

The theoretical and experimental interest in the decays $\eta^{(')}\to\pi^+\pi^-\gamma^{(\ast)}$ encompasses several aspects (see, e.g., Ref.~\cite{gan2020precision} for a review). Firstly, the decays receive contributions from the chiral box anomaly of quantum chromodynamics (QCD) \cite{Wess:1971yu, Witten:1983tw} and allow us to study the $\rho$-$\omega$ mixing mechanism in terms of internal resonance contributions. Secondly, they can be used in a dispersion-theoretical extraction of the form factors for the two-photon interactions of the light pseudoscalar mesons ($\pi^0$, $\eta$, $\eta'$) \cite{Stollenwerk:2011zz, Kubis:2015sga, Hanhart:2016pcd}, which enter the calculation of the hadronic light-by-light (HLbL) scattering contributing to the anomalous magnetic moment of the muon \cite{Blum2016, Jegerlehner:2017gek, Morte_2017}. Furthermore, the decays $\eta^{(')}\to\pi^+\pi^-\gamma^{(\ast)}$ provide a test of P- and CP-violation \cite{Herczeg:1974ik, GENG_2002, GAO_2002} as well as facilitate a search for beyond standard model physics, namely, the search for axion-like particles \cite{gan2020precision}.

In addition to this phenomenological relevance, the decays $\eta^{(')}\to\pi^+\pi^-\gamma^{(\ast)}$ allow us to investigate the symmetry-breaking mechanisms in QCD. In the low-energy regime of QCD, an interplay occurs between dynamical (spontaneous) breaking of chiral symmetry, the explicit symmetry breaking by the quark masses, and the axial $\text{U}(1)_A$ anomaly. For vanishing up-, down-, and strange-quark masses, the QCD Lagrangian at the classical level exhibits a global $\text{U}(3)_L\times \text{U}(3)_R$ chiral symmetry, which is dynamically broken down to $\text{SU}(3)_V\times \text{U}(1)_V$ in the ground state (see, e.g., Ref.~\cite{Scherer:2012xha}). One would then expect the appearance of nine massless pseudoscalar Goldstone bosons \cite{Goldstone:1962}. However, quantum corrections destroy the $\text{U}(1)_V$ symmetry and the singlet axial-vector current is no longer conserved [$\text{U}(1)_A$ anomaly]. As a result, the corresponding singlet Goldstone boson acquires a mass in the chiral limit, as well \cite{tHooft:1976, WITTEN1979269, VENEZIANO1979213}. In the large-number-of-colors (L$N_c$) of QCD \cite{HOOFT1974461, WITTEN197957}, i.e., $N_c\to\infty$  with $g^2 N_c$ fixed, the divergence of the anomalous singlet axial-vector current vanishes, and the singlet pseudoscalar becomes a Goldstone boson in the combined chiral and L$N_c$ limits. In total, this leads to a pseudoscalar nonet ($\pi$, $K$, $\eta_8$, $\eta_1$) as the Goldstone bosons \cite{WITTEN1979269, Coleman:1980}. Therefore, we use massless L$N_c$ QCD as a starting point for perturbative calculations and treat the symmetry breaking by the $\text{U}(1)_A$ anomaly and the nonzero quark masses as corrections. 

At leading order, the decays $\eta^{(')}\to\pi^+\pi^-\gamma^{(\ast)}$ are determined by the chiral box anomaly, which is contained in the Wess-Zumino-Witten (WZW) effective action \cite{Wess:1971yu, Witten:1983tw}. Corrections to the WZW prediction result from the axial $\text{U}(1)_A$ anomaly and the nonzero quark masses. These mechanisms do not only generate masses for the Goldstone bosons but are also responsible for the $\eta-\eta'$ mixing. These effects can be systematically calculated in the framework of large-$N_c$ chiral perturbation theory (L$N_c$ChPT) \cite{LEUTWYLER1996163, HERRERASIKLODY1997345, Kaiser:2000gs}, which is an extension of conventional ChPT \cite{GASSER1985465}, where the pseudoscalar singlet is included. The most general effective Lagrangian of L$N_c$ChPT is organized in a combined expansion in momenta (derivatives), quark masses, and $1/N_c$. Observables are calculated perturbatively, with a power counting determined by a collective small expansion parameter $\delta$ \cite{LEUTWYLER1996163}. 

In this work, we investigate the decays $\eta^{(')}\to\pi^+\pi^-\gamma^{(\ast)}$ at next-to-next-to-leading order (NNLO) in L$N_c$ChPT. Since the dynamical range of the decay involving a real photon, $4M^2_\pi\leq s_{\pi\pi}\leq M^2_{\eta(')}$, is far from the chiral limit, higher-order corrections become important motivating an investigation of their influence. In Sec.~\ref{sec:Lagrangians}, we specify the effective theory we use for our calculations by briefly describing the Lagrangians and the power counting. The calculation of the invariant amplitude is explained in Sec.~\ref{sec:CalcM}, including the $\eta-\eta'$ mixing. Section \ref{sec:etapipigammaNA} contains the numerical evaluation of the results at LO, NLO, and NNLO. In Sec.~\ref{sec:etapipill}, the decays $\eta^{(')}\to\pi^+\pi^-l^+l^-$ involving a lepton pair are discussed, and we conclude with a summary and an outlook of future work in Sec.~\ref{sec:Conclusions}.

\section{Lagrangians and power counting}
\label{sec:Lagrangians}

In the framework of L$N_c$ChPT, we perform a simultaneous expansion of (renormalized) Feynman diagrams
in terms of momenta $p$, quark masses $m$, and $1/N_c$.\footnote{It
	is understood that dimensionful variables need to be small in
	comparison with an energy scale.}
We introduce a collective expansion parameter $\delta$ and count the variables as small quantities of the order of
\cite{Leutwyler:1996sa}
\begin{equation}
\label{powerexp}
p=\mathcal{O}(\sqrt{\delta}),\ \ \ m=\mathcal{O}(\delta),\ \ \
1/N_c=\mathcal{O}(\delta).
\end{equation}
The most general Lagrangian of L$N_c$ChPT is organized as an
infinite series in terms of derivatives, quark-mass terms, and,
implicitly, powers of $1/N_c$, with the scaling behavior given in
Eq.~(\ref{powerexp}):
\begin{equation}
\label{Leff}
\mathcal{L}_{\text{eff}}=\mathcal{L}^{(0)}+\mathcal{L}^{(1)}+\mathcal{L}^{(2)}+\mathcal{L}^{(3)}+\dots,
\end{equation}
where the superscripts $(i)$ denote the order in $\delta$.
In Ref.~\cite{bickert2020}, we explain the power counting and present the relevant Lagrangians, which are used in this work. Here, we briefly outline our approach and refer the reader to Ref.~\cite{bickert2020} for further details.

At leading order, the decays $\eta^{(')}\to\pi^+\pi^-\gamma^{(\ast)}$ are driven by the chiral anomaly in terms of the Wess-Zumino-Witten (WZW) action \cite{Wess:1971yu, Witten:1983tw}, which belongs to the odd-intrinsic-parity sector of the effective field theory.
Since our goal is a one-loop calculation of the decays $\eta^{(')}\to\pi^+\pi^-\gamma^{(\ast)}$, which is NNLO in the $\delta$ counting, we employ the LO, NLO, and NNLO Lagrangians of even intrinsic parity as given in Ref.~\cite{bickert2020}. 
In addition to the WZW action, which starts contributing at $\mathcal{O}(\delta)$, we need the NLO and NNLO Lagrangians from the odd-intrinsic-parity sector, resulting in 
\begin{align}
\mathcal{L}_\epsilon=\mathcal{L}^{(1)}_{\text{WZW}}+\mathcal{L}^{(2)}_\epsilon+\mathcal{L}^{(3)}_\epsilon,
\end{align}
where the superscripts $(i)$ refer to the order in $\delta$. Again, the explicit expressions for the Lagrangians are displayed in Ref.~\cite{bickert2020}. Only the terms of the $\mathcal{O}(p^6)$ Lagrangian need to be updated to those which are specific for the processes of this work. We present them in Table \ref{tab:TermsNcp6} in terms of the building blocks provided in Ref.~\cite{bickert2020}.

\begin{table}[htbp]
	\hspace{-0.75cm}
	\begin{tabular}{c c c c c}\hline\hline
		Lagrangian& LEC& Operator &SU(3)\\
		\hline
			$\mathcal{L}^{(2,N_c p^6)}_\epsilon$	&$L^{6,\epsilon}_1 $&$\langle (\chi)_+ \{ (H_{\mu\nu})_+ (D_\alpha U)_- (D_\beta U)_- + \mbox{rev} \} \rangle\epsilon^{\mu\nu\alpha\beta}$  & x\\
		& $L^{6,\epsilon}_5 $&$\langle (\chi)_- \{ (G_{\mu\nu})_+ (D_\alpha U)_- (D_\beta U)_- -  \mbox{rev} \} \rangle\epsilon^{\mu\nu\alpha\beta}$ & x\\
		& $L^{6,\epsilon}_6 $&$\langle (\chi)_- (D_\mu U)_- (G_{\nu\alpha})_+ (D_\beta U)_-  \rangle\epsilon^{\mu\nu\alpha\beta}$ & x\\
		& $L^{6,\epsilon}_{13} $&$\langle (G_{\mu\nu})_+ \{ (D^\lambda D_\alpha U)^s_- (D_\beta U)_- (D_\lambda U)_- - \mbox{rev} \} \rangle\epsilon^{\mu\nu\alpha\beta}$ & x\\
		& $L^{6,\epsilon}_{14} $&$\langle (G_{\mu\nu})_+ \{ (D_\lambda D_\alpha U)^s_- (D^\lambda U)_- (D_\beta U)_- - \mbox{rev} \} \rangle\epsilon^{\mu\nu\alpha\beta}$ & x\\
		\hline
		$\mathcal{L}^{(3,p^6)}_\epsilon$		& $L^{6,\epsilon}_2 $&$\langle (\chi)_+ (D_\mu U)_- \rangle\langle (D_\nu U)_- (H_{\alpha\beta})_+ \rangle\epsilon^{\mu\nu\alpha\beta}$ & x\\
		&$L^{6,\epsilon}_7 $&$\langle (\chi)_- \rangle\langle (G_{\mu\nu})_+ (D_\alpha U)_- (D_\beta U)_- \rangle\epsilon^{\mu\nu\alpha\beta}$& x\\
		&$L_{227}$&$i \epsilon^{\mu\nu\lambda\rho}\la\nabla^{\sigma}f_{+\mu\sigma}u_{\nu}u_{\lambda}\ra\la u_{\rho}\ra$& $\cdots$\\
		&$L_{228}$&$i \epsilon^{\mu\nu\lambda\rho}\la\nabla^{\sigma}f_{+\mu\nu}u_{\sigma}u_{\lambda}\ra\la u_{\rho}\ra+\mbox{H.c.}$& $\cdots$\\
		&$L_{229}$&$i \epsilon^{\mu\nu\lambda\rho}\la{f_{+\mu}}^{\sigma}h_{\nu\sigma}u_{\lambda}\ra\la u_{\rho}\ra+\mbox{H.c.}$& $\cdots$\\
		&$L_{230}$&$i \epsilon^{\mu\nu\lambda\rho}\la f_{+\mu\nu}{h_{\lambda}}^{\sigma}u_{\sigma}\ra\la u_{\rho}\ra+\mbox{H.c.}$& $\cdots$\\
		&$L_{233}$&$i \epsilon^{\mu\nu\lambda\rho}\la\nabla^{\sigma}f_{+\mu\sigma}\ra\la u_{\nu}u_{\lambda}u_{\rho}\ra$& $\cdots$\\
		&$L_{234}$&$i \epsilon^{\mu\nu\lambda\rho}\la{f_{+\mu}}^{\sigma}\ra\la u_{\nu}u_{\lambda}h_{\rho\sigma}\ra$& $\cdots$\\
		&$L_{242}$&$\epsilon^{\mu\nu\lambda\rho}\la u_{\mu}\ra\la u_{\nu}f_{-\lambda\rho}\chip\ra
		+\mbox{H.c.}$& $\cdots$\\
		&$L_{254}$&$\epsilon^{\mu\nu\lambda\rho}\la f_{+\mu\nu}\ra\la u_{\lambda}u_{\rho}\chim\ra$& $\cdots$\\
		&$L_{255}$&$\epsilon^{\mu\nu\lambda\rho}\la f_{+\mu\nu}\chim u_{\lambda}\ra\la u_{\rho}\ra
		+\mbox{H.c.}$& $\cdots$\\
		&$\Lambda_{437}$&$(\psi+\theta)\left(i \epsilon^{\mu\nu\lambda\rho}\la f_{+\mu\nu}\chip u_{\lambda}u_{\rho}\ra
		+\mbox{H.c.}\right)$& $\cdots$\\
		&$\Lambda_{438}$&$ i \epsilon^{\mu\nu\lambda\rho}(\psi+\theta)\la f_{+\mu\nu}u_{\lambda}\chip u_{\rho}\ra$& $\cdots$\\
		\hline\hline
	\end{tabular}
	\caption{Relevant terms of $\mathcal{L}^{(2,N_c p^6)}_\epsilon$ and $\mathcal{L}^{(3,p^6)}_\epsilon$.}
	\label{tab:TermsNcp6}
\end{table}

\section{Calculation of the invariant amplitude}\label{sec:CalcM}

\begin{figure}
	\centering
	\includegraphics[width=0.9\textwidth]{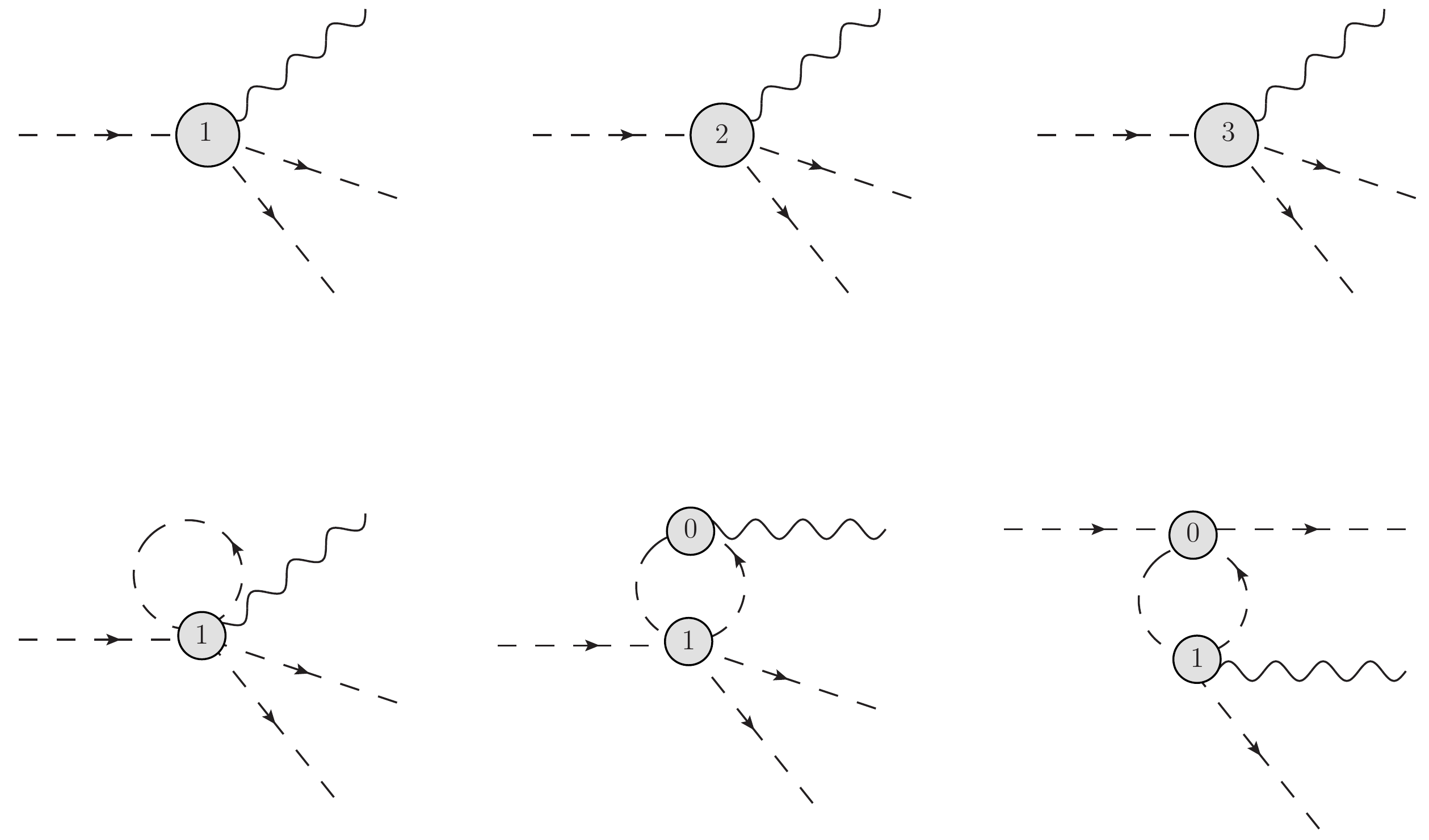}
	\caption{Feynman diagrams for $\eta^{(')}\to\pi^+\pi^-\gamma^*$ up to and including NNLO. Dashed lines refer to pseudoscalar mesons and wiggly lines to photons.  The numbers $k$ in the interaction blobs refer to vertices derived from the corresponding Lagrangians $\mathcal{L}^{(k)}$.}
	\label{fig:phigallp}
\end{figure}

The invariant amplitude for the decay $P\to\pi^+\pi^-\gamma^{(\ast)}$ of a pseudoscalar meson $P$ can be parameterized by 
\begin{align}
	\mathcal{M}=-iF_P\epsilon_{\mu\nu\alpha\beta}\epsilon^\mu p^\nu_1 p^\alpha_2 q^\beta,
\end{align}
where $q^\mu$ and $\epsilon^\mu$ denote the momentum and polarization vector of the photon, respectively, and $p^\mu_1$, $p^\mu_2$ are the momenta of the pions with $s_{\pi\pi}=(p_1+p_2)^2$. To obtain the invariant amplitude up to and including NNLO, we have to evaluate the Feynman diagrams shown in Fig.~\ref{fig:phigallp}, where the
vertices are obtained from the Lagrangians given in Sec.~\ref{sec:Lagrangians} and in Ref.~\cite{bickert2020}.

The coupling to the electromagnetic field is described by introducing an external field which couples to the electromagnetic current operator
\begin{align}
J^\mu=\bar{q}Q\gamma^\mu q,
\end{align}
where $Q$ is the quark-charge matrix. For $N_c=3$, the quark-charge matrix is given by
\begin{align}
Q(3)=\text{diag}\left(\frac{2}{3},-\frac{1}{3},-\frac{1}{3}\right).
\end{align}
However, B\"ar and Wiese pointed out \cite{Bar:2001qk} that in order for the Standard Model to be consistent for arbitrary $N_c$, the ordinary quark-charge matrix should be replaced by (see also Ref.~\cite{Gerard:1995bv})
\begin{align}
Q(N_c)=\frac{1}{2}\text{diag}\left(\frac{1}{N_c}+1,\frac{1}{N_c}-1,\frac{1}{N_c}-1\right)
=-\frac{1}{6}{\mathbbm 1}+\frac{1}{2}\lambda_3+\frac{1}{2\sqrt{3}}\lambda_8
+\frac{1}{2N_c}{\mathbbm 1}.
\end{align}
Therefore, we use $Q(N_c)$ for the calculation of the invariant amplitude. However, in the evaluation of the Feynman diagrams, it turns out  that, due to the flavor structure, the $N_c$-dependent part of $Q(N_c)$ gives no contribution to the matrix element. 
The Feynman diagrams are calculated using the \textit{Mathematica} package FEYNCALC \cite{Mertig:1990an}.

Furthermore, we take into account the $\eta$-$\eta'$ mixing at NNLO, following the detailed derivation of the mixing in Ref.~\cite{Bickert:2016fgy}. We start by calculating the coupling of the pions and the photon to the octet and singlet fields $\phi_b$, collected in the doublet $\eta_A\equiv\left(\eta_8,\eta_1\right)^T$, at the one-loop level up to and including NNLO in the $\delta$ counting. The result, which should be interpreted as a Feynman rule, is given by the ``matrix elements'' $\mathcal{F}_{b}=\left\langle \pi^+\pi^-\gamma^*|b\right\rangle$. Then, we transform the bare fields $\eta_A$ to the physical states using the transformation $T$ in Eq.~(51) in Ref.~\cite{Bickert:2016fgy}:
\begin{align}
\begin{pmatrix} \eta_8 \\
\eta_1
\end{pmatrix}=
\begin{pmatrix} T_{8\eta} & T_{8\eta'}\\
T_{1\eta} & T_{1\eta'}
\end{pmatrix}
\begin{pmatrix} \eta \\
\eta'
\end{pmatrix}.
\end{align}
The resulting (``physical'') matrix elements are then obtained from
\begin{align}
\begin{pmatrix} F_{\eta} \\
F_{\eta'}
\end{pmatrix}=
\begin{pmatrix} T_{8\eta} & T_{1\eta}\\
T_{8\eta'} & T_{1\eta'}
\end{pmatrix}
\begin{pmatrix} \mathcal{F}_8 \\
\mathcal{F}_1
\end{pmatrix}.
\end{align}
For the calculation of the loop diagrams, we employ the LO mixing.

At LO and NLO, the form factors $F_P$ are given by
\begin{align}
	F^\text{LO}_\eta
	&=\frac{1}{4 \sqrt{3} \pi ^2 F_{\pi }^3}\left[\cos (\theta^{[0]} )-\sqrt{2} \sin (\theta^{[0]} )\right],\label{eq:FFetaLO}\\
	F^\text{LO}_{\eta'}
	&=\frac{1}{4 \sqrt{3} \pi ^2 F_{\pi }^3}\left[\sin (\theta^{[0]}
	)+\sqrt{2} \cos (\theta^{[0]} )\right],
	\label{eq:FFetapLO}
\end{align}
\begin{align}
	F^\text{NLO}_\eta
	&=\frac{1}{4 \sqrt{3} \pi ^2 F_{\pi }^3}\left\{\left[\cos (\theta^{[1]} )-\sqrt{2} \sin (\theta^{[1]} )\right] \left(1+c_{14} M_{\eta }^2+c_{13}
	M_{\pi }^2-c_{14} q^2+c_{15} s_{\pi\pi}\right)\right.\no\\
	&\quad\left.-\sqrt{2} \sin (\theta^{[1]} )c_2\right\},\label{eq:FFeta}\\
	F^\text{NLO}_{\eta'}
	&=\frac{1}{4 \sqrt{3} \pi ^2 F_{\pi }^3}\left\{\left[\sin (\theta^{[1]}
	)+\sqrt{2} \cos (\theta^{[1]} )\right] \left(1+c_{14} M_{\text{$\eta'$}}^2+c_{13} M_{\pi }^2-c_{14}
	q^2+c_{15} s_{\pi\pi}\right)\right.\no\\
	&\quad\left.+\sqrt{2} \cos (\theta^{[1]} )c_2\right\},
	\label{eq:FFetap}
\end{align}
where
\begin{align}
	c_2&=-48\pi^2 \tilde{L}_1-\frac{\Lambda_1}{2},\no\\
	c_{13}&=-1024\pi^2\left(L^{6,\epsilon}_{13}+L^{6,\epsilon}_{14}+L^{6,\epsilon}_{5}+\frac{L^{6,\epsilon}_{6}}{2}\right),\no\\
	c_{14}&=512\pi^2L^{6,\epsilon}_{13},\no\\
	c_{15}&=512\pi^2\left(2L^{6,\epsilon}_{13}+L^{6,\epsilon}_{14}\right),
\end{align}
and $\theta^{[i]}$ is the corresponding mixing angle at LO(NLO), given in Eq.~(49) in Ref.~\cite{Bickert:2016fgy}. The parameter $c_2$ represents a QCD-scale-invariant combination of parameters violating the Okubo-Zweig-Iizuka (OZI) rule \cite{Kaiser:2000gs}.
Since the expressions at NNLO are very long, we only display the loop corrections, corresponding to the loop diagrams in Fig.~\ref{fig:phigallp}, in Appendix \ref{app:Expressions}. However, the tree-level contributions can be provided as a \textit{Mathematica} notebook.
At NNLO, we have to deal with a proliferation of LECs and the fact that the $\mathcal{O}(p^8)$ Lagrangian, which should be taken into account according to our power counting, has not been constructed. Therefore, we make the following ansatz for the form factors at NNLO:
\begin{align}
	F^\text{NNLO}_\eta(s_{\pi\pi})&=F^{\text{LO}}_\eta+\frac{1}{4\sqrt{3}\pi^2F^3_\pi}\left(b_\eta+c_{\eta} s_{\pi\pi}+d_{\eta}s^2_{\pi\pi}\right)+\text{loops}_\eta(s_{\pi\pi}),\label{ansatzEta}\\
	F^\text{NNLO}_{\eta'}(s_{\pi\pi})&=F^{\text{LO}}_{\eta'}+\frac{1}{4\sqrt{3}\pi^2F^3_\pi}\left(b_{\eta'}+c_{\eta'} s_{\pi\pi}+d_{\eta'}s^2_{\pi\pi}\right)+\text{loops}_{\eta'}(s_{\pi\pi})\label{ansatzEtap},
\end{align}
where $F^{\text{LO}}_P$ are the LO form factors given in Eqs.~(\ref{eq:FFetaLO}) and (\ref{eq:FFetapLO}), and the expression $\text{loops}_P(s_{\pi\pi})$ refers to the $s_{\pi\pi}$-dependent parts of the loop corrections. 
The parameters $b_P$ and $c_{P}$ receive contributions from the higher-order Lagrangians in Sec.~\ref{sec:Lagrangians} and Ref.~\cite{bickert2020} as well as from, in principle, the $\mathcal{O}(p^8)$ Lagrangian. In addition, the LECs and loop contributions originating from the $\eta$-$\eta'$ mixing are also absorbed in $b_P$ and $c_{P}$. The parameters $d_{P}$ consist solely of terms from the $\mathcal{O}(p^8)$ Lagrangian. 
However, the most general form factor at NNLO could depend on a second kinematic variable $t$ or $u$. This dependence would be introduced by the $\mathcal{O}(p^8)$ Lagrangian. For simplicity, we ignore those contributions in the following and employ the ansatz in Eqs.~(\ref{ansatzEta}) and (\ref{ansatzEtap}). 

A measurable observable of the decay is provided by the differential cross section as a function of the photon energy 
\begin{align}
	\omega=\frac{1}{2}\left(M_P-\frac{s_{\pi\pi}}{M_P}\right), 
\end{align}
which takes the form \cite{Hacker:2008}
\begin{align}
	\frac{d\Gamma}{d\omega}=\frac{M_P \omega^3(M^2_P-4M^2_\pi-2M_P\omega)}{384\pi^3}\sqrt{1-\frac{4M^2_\pi}{M^2_P-2M_P\omega}}\left|F_P\right|^2.
	\label{eq:spectrum}
\end{align}
The full decay width can then be obtained by integration
\begin{align}
	\Gamma_{P\to\pi^+\pi^-\gamma}=\int^{\frac{1}{2}(M_P-4M^2_\pi/M_P)}_{0}d\omega \frac{d\Gamma}{d\omega}.
	\label{eq:FullGamma}
\end{align}

\section{Numerical analysis}\label{sec:etapipigammaNA}

To evaluate our results numerically we need to fix the LECs. This is done in a successive way, starting at LO and proceeding to NLO and, finally, to NNLO.

\subsection{LO}
\label{sec:SpectraLO}

At LO, we can directly calculate the decay widths by using Eq.~(\ref{eq:FullGamma}) together with the form factors in Eqs.~(\ref{eq:FFetaLO}) and (\ref{eq:FFetapLO}). The LO results are
\begin{align}
	\Gamma_{\eta\to\pi^+\pi^-\gamma}&=36\ \text{eV},\\
	\Gamma_{\eta'\to\pi^+\pi^-\gamma}&=3.4\ \text{keV},
\end{align}
which, in particular for the $\eta'$, are a lot smaller than the experimental values $\Gamma_{\eta\to\pi^+\pi^-\gamma}=(55.3\pm 2.4)\ \text{eV}$ \cite{Zyla:2020zbs} and $\Gamma_{\eta'\to\pi^+\pi^-\gamma}=(55.5\pm 1.9)\ \text{keV}$ \cite{Zyla:2020zbs}.
Employing Eq.~(\ref{eq:spectrum}) with the LO form factors, we also determine the spectra at LO and compare them to the experimental data. Since the data are provided in arbitrary units, we multiply our LO results for the spectra by a normalization constant $A_P$, $P=\eta,\ \eta'$, and determine this constant through a fit to the data. For $\eta\to\pi^+\pi^-\gamma$ we use the full photon-energy spectrum provided by Ref.~\cite{Adlarson:2011xb}, and for $\eta'\to\pi^+\pi^-\gamma$ we fit our results to the $\pi^+\pi^-$ invariant-mass spectrum, measured in Ref.~\cite{Ablikim:2017fll}, up to $0.59$ GeV. 
The results are shown in Fig.~\ref{fig:nlopl}. As one can clearly see, the LO description is very poor and it is crucial to take higher-order corrections into account.

\subsection{NLO}
\label{sec:pipigammaNLO}

At NLO, we determine the appearing LECs through a fit to the experimental spectra of the decays. It is not possible to independently determine all NLO LECs in the expressions for the NLO form factors in Eqs.~(\ref{eq:FFeta}) and (\ref{eq:FFetap}). We are only able to fix those linear combinations of LECs which accompany independent $s_{\pi\pi}$ structures. The NLO form factors in terms of these linear combinations of LECs are given by
\begin{align}
	F_\eta(s_{\pi\pi})
	&=\frac{1}{4 \sqrt{3} \pi ^2 F_{\pi }^3}\left\{\left[\cos (\theta^{[1]} )-\sqrt{2} \sin (\theta^{[1]} )\right]\left(1+c_{15}s_{\pi\pi}\right)+c_3\right\},\label{eq:fitNLO}\\
	F_{\eta'}(s_{\pi\pi})&=
	\frac{1}{4 \sqrt{3} \pi ^2 F_{\pi }^3}\left\{\left[\sin (\theta^{[1]}
	)+\sqrt{2} \cos (\theta^{[1]} )\right]\left(1+c_{15}s_{\pi\pi}\right)+c_4\right\},
	\label{eq:fitNLOp}
\end{align}
where $\theta^{[1]}$ is the mixing angle calculated up to and including NLO, given in Eq.~(49) in Ref.~\cite{Bickert:2016fgy}, and
\begin{align}
	c_3&=\left[\cos (\theta^{[1]})-\sqrt{2}\sin(\theta^{[1]})\right]\left(c_{13}M^2_\pi+c_{14}M^2_\eta\right)-\sqrt{2}\sin(\theta^{[1]})\, c_2,\no\\
	c_4&=\left[\sin(\theta^{[1]})+\sqrt{2}\cos(\theta^{[1]})\right]\left(c_{13}M^2_\pi+c_{14}M^2_{\eta'}\right)+\sqrt{2}\cos(\theta^{[1]})\, c_2.
\end{align}
We now have to determine four parameters $c_3$, $c_4$, $c_{15}$, and the NLO mixing angle $\theta^{[1]}$. For $\theta^{[1]}$ we employ the value from the NLO analysis in Table IV in Ref.~\cite{Bickert:2016fgy}, labeled NLO 1, namely $\theta^{[1]}=-11.1\deg$. The constants $c_3$, $c_4$, and $c_{15}$ are determined through a fit to experimental data. We use the decay width of $\eta\to\pi^+\pi^-\gamma$, the photon-energy spectrum of the $\eta$ decay, and the $\pi^+\pi^-$ invariant-mass spectrum of the $\eta'$ decay. Since we are not able to describe the full $\eta'$ spectrum, we do not include the $\eta'$ decay width in our fit. We perform three simultaneous fits to the data for the $\eta$ decay width \cite{Zyla:2020zbs}, the full $\eta$ spectrum from Ref.~\cite{Adlarson:2011xb}, and to the $\eta'$ spectrum from Ref.~\cite{Ablikim:2017fll} up to $0.59$ GeV (I), $0.64$ GeV (II), and $0.72$ GeV (III). 
Since the experimental spectra are provided in arbitrary units, we multiply our fit functions, i.e., Eq.~(\ref{eq:spectrum}) with the form factors from Eqs.~(\ref{eq:fitNLO}) and (\ref{eq:fitNLOp}), by normalization constants $A_P$. The results for the fit parameters are given in Table \ref{tab:FitParametersAtNLO}, where the errors are the ones provided by the \textit{Mathematica} fit routine NonlinearModelFit. In all fits, in the calculation of the fit parameter errors, we only take the experimental errors into account. To that end, the estimated variance, corresponding to the reduced $\chi^2$, is set to 1. In order to evaluate the quality of the fits, we display the mean squared error denoted by MSE in the tables for the fit parameters. The MSE can be obtained from the ANOVATable in \textit{Mathematica} and is defined as
\begin{align}
\text{MSE}=\frac{1}{n_\text{dof}}\sum^{N}_{i=1}\frac{\left(y_i-\hat{y_i}\right)^2}{\Delta y_i^2},
\end{align}
where $n_\text{dof}$ is the number of degrees of freedom, $N$ the number of data points, $y_i$ the value of the $i$th data point, $\Delta y_i$ its error, and $\hat{y_i}$ the corresponding model prediction.
Furthermore, we do not consider the errors caused by neglecting higher-order terms. In principle, a systematic error of at least 10\%, corresponding to $\delta^2=1/9$, should be added to all quantities determined up to and including NLO.

\begin{table}[htbp]
	\centering
	\begin{tabular}{c  r@{$\,\pm\,$}l r@{$\,\pm\,$}l r@{$\,\pm\,$}l r@{$\,\pm\,$}l r@{$\,\pm\,$}l c}\myTR
		Fit	& \multicolumn{2}{c}{$A_\eta$ $[10^{10}]$} & \multicolumn{2}{c}{$A_{\eta'}$  $[10^{8}]$} & \multicolumn{2}{c}{$c_3$} & \multicolumn{2}{c}{$c_4$} & \multicolumn{2}{c}{$c_{15}\ [\text{GeV}^{-2}]$} & MSE \\ \myMR
\text{I} & $1.43$&$0.06$ & $-0.85$&$0.07$ & $-0.68$&$0.04$ & $-0.86$&$0.02$ & $5.78$&$0.23$ & 7.58 \\
\text{II} & $1.43$&$0.06$ & $-1.32$&$0.11$ & $-0.68$&$0.04$ & $-1.24$&$0.01$ & $5.78$&$0.23$ & 29.89 \\
\text{III} & $1.43$&$0.06$ & $-3.06$&$0.25$ & $-0.68$&$0.04$ & $-1.89$&$0.03$ & $5.78$&$0.23$ & 221.96 \\
		\myBR
	\end{tabular}
	\caption{Fit parameters at NLO.}
	\label{tab:FitParametersAtNLO}
\end{table}

The parameters $A_{\eta}$ and $c_3$ appear only in the $\eta$ form factor and are therefore fixed by the $\eta$ data. Because the fit range of these data remains the same in the three cases, the parameters do not change. Also $c_{15}$, which appears in both the expression for the $\eta$ and the $\eta'$ form factor, seems to be determined by the $\eta$ spectrum, since it does not depend on the fit range of the $\eta'$ spectrum. The variation of the $\eta'$ fit range is then reflected in the variation of $A_{\eta'}$ and $c_4$. A vector-meson-dominance (VMD) estimate from SU(3) ChPT predicts $c_{15}=2.53\ \text{GeV}^{-2}$ \cite{Hacker:2008}. Our value for $c_{15}$ is more than twice as large.

\begin{figure}[htbp]
	\includegraphics[width=1.0\textwidth]{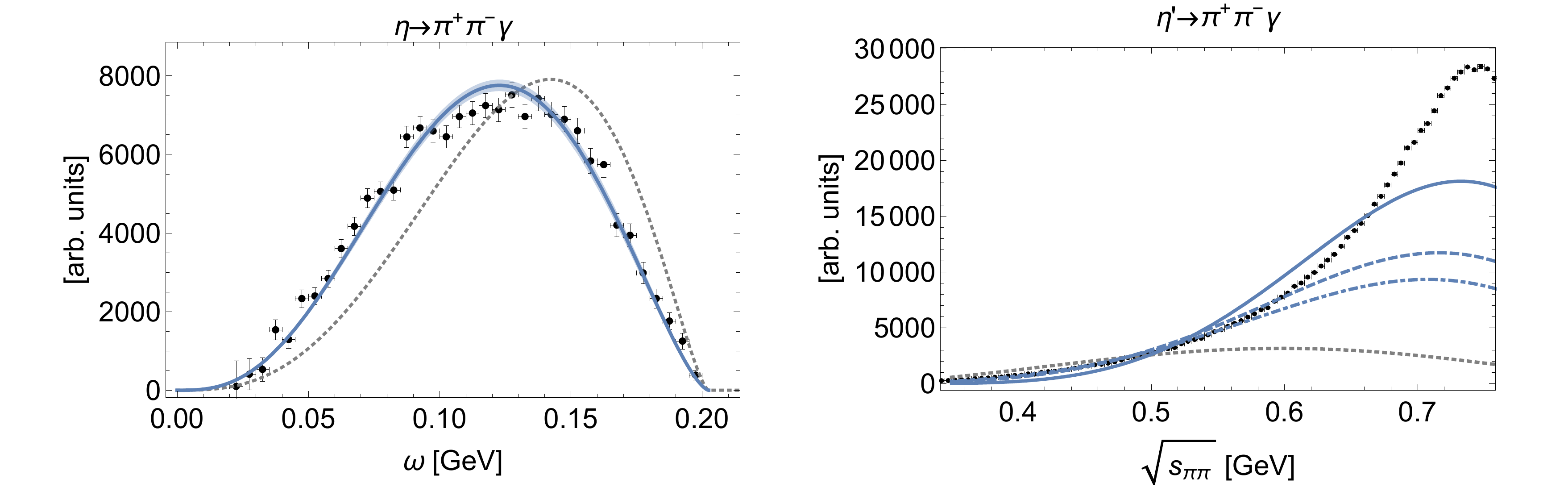}
	\caption{Left: Photon-energy spectrum of $\eta\to\pi^+\pi^-\gamma$ at LO (dotted, gray) and NLO (solid, blue). The blue band is the $1\sigma$ error band. The experimental data are taken from Ref.~\cite{Adlarson:2011xb}. Right: Invariant-mass spectrum of the $\pi^+\pi^-$ system in $\eta'\to\pi^+\pi^-\gamma$ at LO (dotted, gray) and NLO (blue) fitted up to $0.59$ GeV (dash-dotted), $0.64$ GeV (dashed), $0.72$ GeV (solid). The experimental data are taken from Ref.~\cite{Ablikim:2017fll}.}
	\label{fig:nlopl}
\end{figure}

\begin{figure}[htbp]
	\centering
	\includegraphics[width=1.0\textwidth]{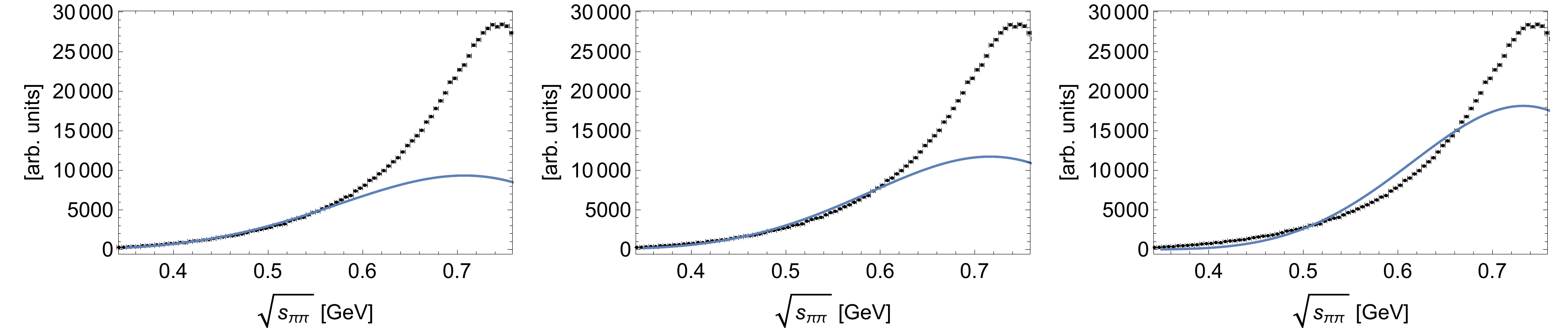}
	\caption{Invariant-mass spectrum of the $\pi^+\pi^-$ system in $\eta'\to\pi^+\pi^-\gamma$ with the 1$\sigma$ error band, coinciding with the line thickness, at NLO fitted up to $0.59$ GeV (left), $0.64$ GeV (middle), $0.72$ GeV (right). The experimental data are taken from Ref.~\cite{Ablikim:2017fll}.}
	\label{fig:bnlop}
\end{figure}

The NLO results for the $\eta$ and $\eta'$ spectra are shown in Fig.~\ref{fig:nlopl} together with the LO results obtained in Sec.~\ref{sec:SpectraLO} and the experimental data. The 1$\sigma$ error bands of the fits of the $\eta'$ spectra are displayed in Fig.~\ref{fig:bnlop}.
For both the $\eta$ and the $\eta'$ spectrum, the NLO description is a clear improvement compared to the LO result. At NLO, increasing the fit range of the $\eta'$ spectrum leads to a better description of the data at higher $s_{\pi\pi}$, but it worsens at lower $s_{\pi\pi}$. The error bands for the $\eta'$ spectra are so small that they coincide with the line thickness. This is caused by the fact that the fit errors are calculated only from the experimental errors which are very small. From our analysis of the $\eta'$ decay we conclude that a NLO calculation should not be applied to data with $\sqrt{s_{\pi\pi}}$ larger than 0.6 GeV, which motivates going to NNLO.

\subsection{NNLO}

At NNLO, we employ the ansatz for the form factors in Eqs.~(\ref{ansatzEta}) and (\ref{ansatzEtap}). Since the form factors for $\eta$ and $\eta'$ each have their specific set of LECs, we perform the fits to the corresponding data separately. The normalization $A_\eta$ and the LECs $b_\eta$, $c_\eta$, $d_\eta$ are fixed through a simultaneous fit to the $\eta$ decay width \cite{Zyla:2020zbs} and the photon-energy spectrum \cite{Adlarson:2011xb}. We consider four different scenarios. The first is the full NNLO calculation (Full). In a next step, we switch off the loop contributions (Without loops). Finally, we put the $d_\eta$ term to zero, and we also discuss the case without $d_\eta$ and without loop contributions. The results are shown in Table \ref{tab:NNLOeta} in Appendix \ref{app:Par}.
Then, all four scenarios are discussed for the $\eta'$. Since we cannot describe the full $\eta'$ spectrum, we do not include the decay width in the fit. As a result, when the loop contributions are switched off, we are not able to extract the overall normalization separately. In those cases, we can only fit the spectrum induced by the form factor
\begin{align}
	F^\text{NNLO}_{\eta'}(s_{\pi\pi})&=F^{\text{LO}}_{\eta'}+\frac{1}{4\sqrt{3}\pi^2F^3_\pi}\left(\tilde{c}_{\eta'} s_{\pi\pi}+\tilde{d}_{\eta'}s^2_{\pi\pi}\right)
\end{align}
multiplied by a normalization constant $\tilde{A}_{\eta'}$. The relation to the parameters given in Eq.~(\ref{ansatzEtap}) (without $\text{loops}_P(s_{\pi\pi})$), with the original normalization $A_{\eta'}$, takes the form
\begin{align}
	\sqrt{\tilde{A}_{\eta'}}&=\frac{\sin(\theta^{[0]})+\sqrt{2}\cos(\theta^{[0]})+b_{\eta'}}{\sin(\theta^{[0]})+\sqrt{2}\cos(\theta^{[0]})}\sqrt{A_{\eta'}},\no\\
	\tilde{c}_{\eta'}&=\frac{\sin(\theta^{[0]})+\sqrt{2}\cos(\theta^{[0]})}{\sin(\theta^{[0]})+\sqrt{2}\cos(\theta^{[0]})+b_{\eta'}}c_{\eta'},\no\\
	\tilde{d}_{\eta'}&=\frac{\sin(\theta^{[0]})+\sqrt{2}\cos(\theta^{[0]})}{\sin(\theta^{[0]})+\sqrt{2}\cos(\theta^{[0]})+b_{\eta'}}d_{\eta'},
\end{align}
where $\theta^{[0]}=-19.6\deg$ is the LO mixing angle.
In the scenarios including loops, the loop contributions provide additional independent $s_{\pi\pi}$ structures, so we can try to extract the LECs and the overall normalization separately. The results with and without loops are provided in Tables \ref{tab:NNLOetapl} and \ref{tab:NNLOetap} in Appendix \ref{app:Par}, respectively.

Figure \ref{fig:bandtog} shows our LO, NLO, and NNLO predictions for the $\eta$ spectrum together with the experimental data.
\begin{figure}[htbp]
	\centering
	\includegraphics[width=0.8\textwidth]{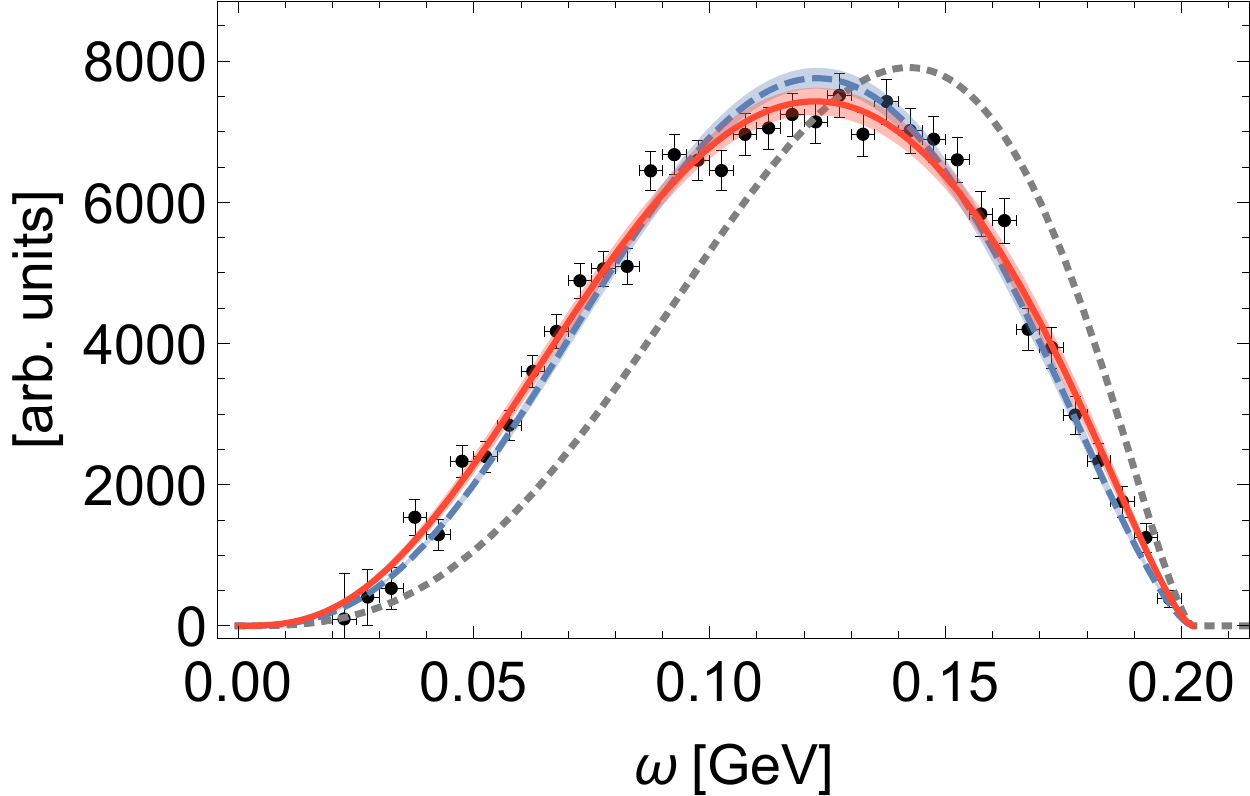}
	\caption{Photon-energy spectrum of $\eta\to\pi^+\pi^-\gamma$ at LO (dotted, gray), NLO (dashed, blue) and NNLO (solid, red). For the NLO and NNLO results the corresponding 1$\sigma$ error bands are shown. The experimental data are taken from Ref.~\cite{Adlarson:2011xb}.}
	\label{fig:bandtog}
\end{figure}
As expected, the description of the spectrum improves gradually from LO to NLO to NNLO. We find that the contributions of the loops to the shape of the spectrum are very small and can be compensated by a change of the LECs. The improved description of the data from NLO to NNLO is due to the inclusion of the $s^2_{\pi\pi}$ term.

\begin{figure}[htbp]
	\centering
	\includegraphics[width=0.80\textwidth]{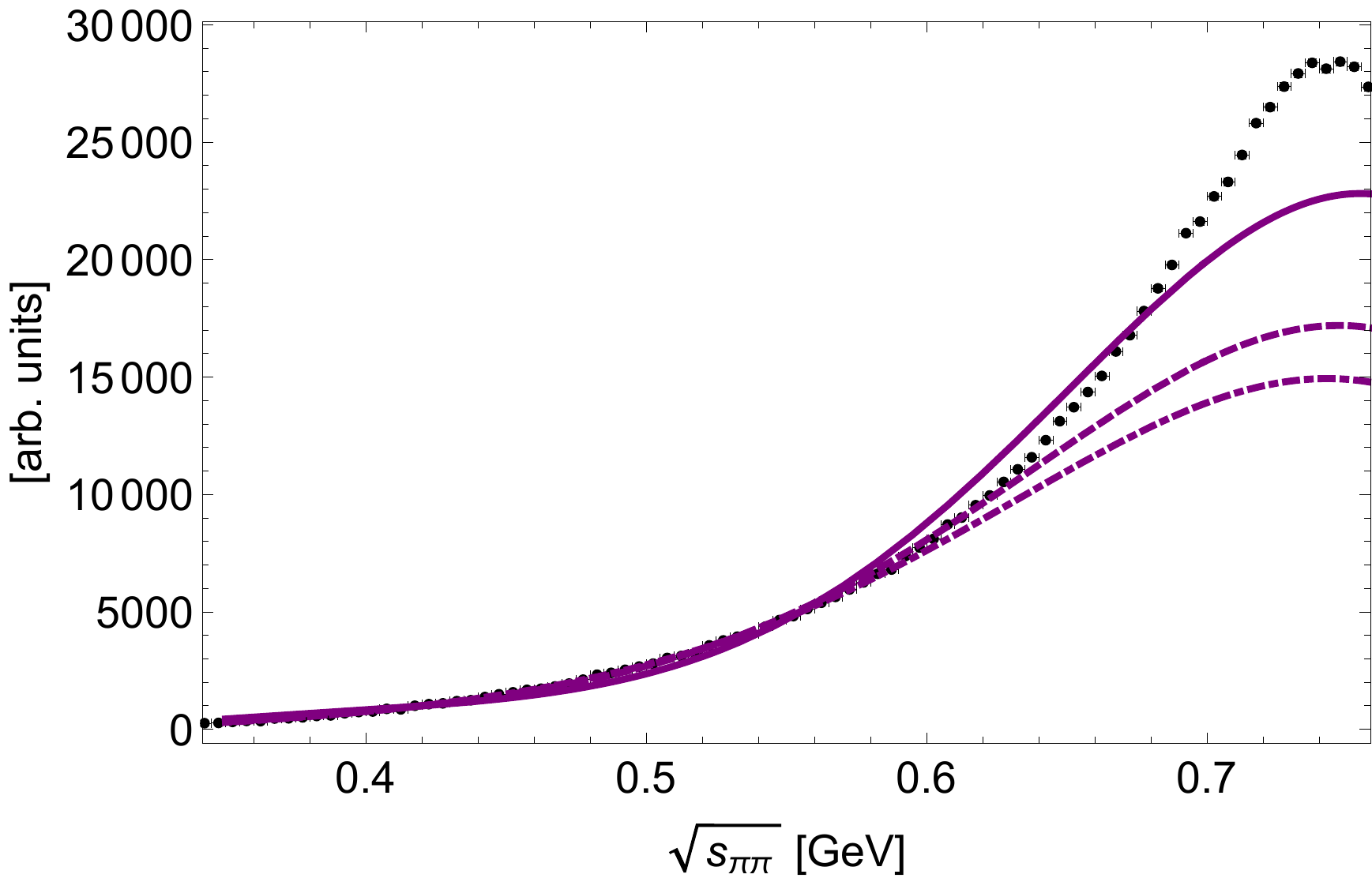}
	\caption{Invariant-mass spectrum of the $\pi^+\pi^-$ system in $\eta'\to\pi^+\pi^-\gamma$ at NNLO with $d_{\eta'}=0$, fitted up to $0.59$ GeV (dash-dotted), $0.64$ GeV (dashed), $0.72$ GeV (solid). The experimental data are taken from Ref.~\cite{Ablikim:2017fll}.}
	\label{fig:plNNLO}
\end{figure}

Figure \ref{fig:plNNLO} shows the results of the fits of the NNLO expression for the $\eta'$ spectrum to the experimental data without the $s^2_{\pi\pi}$ term in the three different fit ranges. The corresponding error bands are displayed in Fig.~\ref{fig:b2p} in Appendix \ref{app:plots}. 
Here, we observe a better description of the data compared to the NLO calculation due to the inclusion of the loop corrections and the appearance of an additional parameter, because the LEC multiplying the $s_{\pi\pi}$ term, i.e. $c_{\eta'}$, is now independent from the $\eta$ decay. Taking the $s^2_{\pi\pi}$ term into account in the full NNLO expression tends to make the fit unstable, in particular in the cases where the fit range is small. Therefore, we discuss here only the results of the fits up to $0.72$ GeV (III) and the results of the other fits are shown in Fig.~\ref{fig:b2p1} in Appendix \ref{app:plots}.
Figure \ref{fig:gridall} shows a comparison of our NLO, NNLO without the $d_{\eta'}$ term, and full NNLO results for the $\eta'$ spectrum fitted up to $0.72$ GeV. 
\begin{figure}[htbp]
	\centering
	\includegraphics[width=1.0\textwidth]{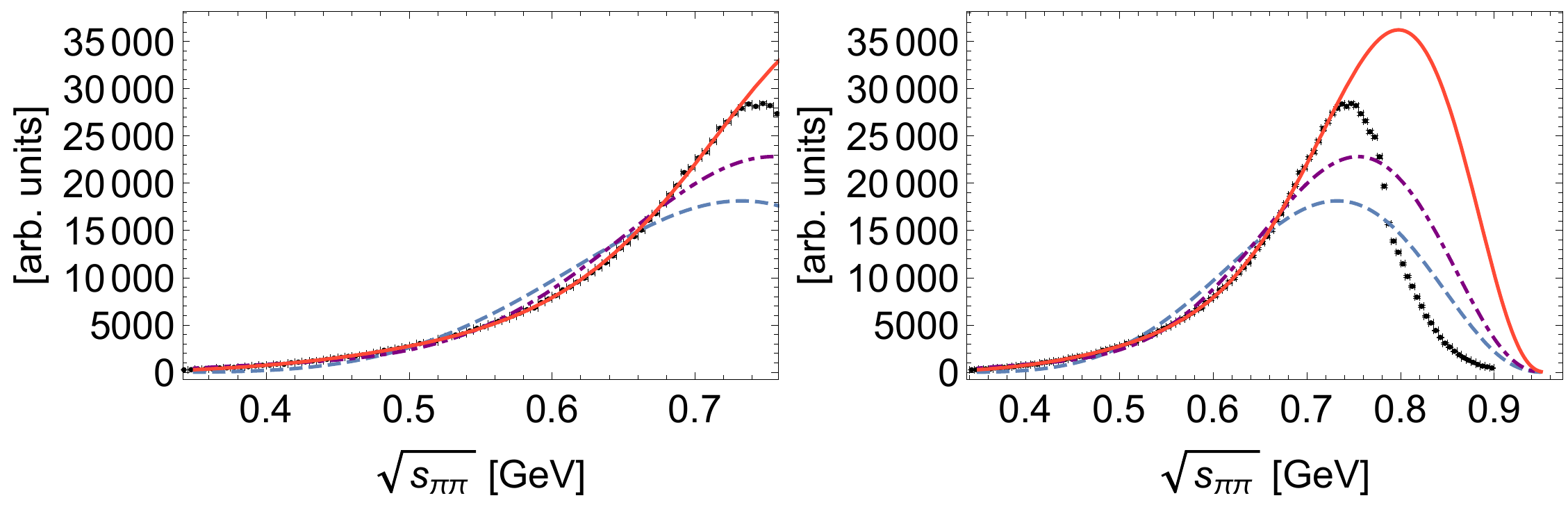}
	\caption{Invariant-mass spectrum of the $\pi^+\pi^-$ system in $\eta'\to\pi^+\pi^-\gamma$ at NLO (dashed, blue), NNLO with $d_{\eta'}=0$ (dash-dotted, purple), and full NNLO (solid, red) fitted up to $0.72$ GeV. The left plot shows the spectrum up to $0.75$ GeV and the right plot the full spectrum. The experimental data are taken from Ref.~\cite{Ablikim:2017fll}.}
	\label{fig:gridall}
\end{figure}
At such high values of $s_{\pi\pi}$, the inclusion of the $d_{\eta'}$ term yields a better description of the data compared to NNLO with $d_{\eta'}=0$.
However, as can be seen in Fig.~\ref{fig:gridall}, even the full NNLO result is not able to describe the whole spectrum. This problem originates from the fact that, since the invariant mass of the pion pair reaches values as high as $0.8$ GeV, vector-meson degrees of freedom become important. Since we do not consider vector mesons as explicit degrees of freedom in our calculation, we cannot reproduce the whole spectrum correctly.

\subsubsection{Comparison with other works}

The decay $\eta\to\pi^+\pi^-\gamma$ has been studied in one-loop ChPT using the LO $\eta$-$\eta'$ mixing in Refs.~\cite{Bijnens:1989ff, Hacker:2008}. It was found that $\mathcal{O}(p^6)$ corrections are crucial to describe the data, and that the contributions of the contact terms dominate over the loop corrections. We agree with these findings. 
Reference \cite{Borasoy:2004qj} investigates the decays $\eta^{(')}\to\pi^+\pi^-\gamma$ in an approach that combines ChPT with a coupled-channel Bethe-Salpeter equation which generates vector mesons dynamically. The importance of $\mathcal{O}(p^6)$ contact terms for describing the data for the $\eta$ decay was also observed. The $\eta'$ data however, cannot be described by simply adjusting the $\mathcal{O}(p^6)$ contact terms. In the decay $\eta'\to\pi^+\pi^-\gamma$, vector mesons play an important role and, after the inclusion of the coupled-channel approach, the experimental $\eta'$ spectrum can be reproduced. 
The effects of vector mesons have been taken into account by a momentum dependent vector-meson-dominance model \cite{Picciotto:1991ae} or, in a more elaborate way, in the context of Hidden Local Symmetries \cite{Benayoun:2003we, Benayoun:2009im}. In Ref.~\cite{Osipov:2020vad}, axial-vector mesons and their mixing with pseudoscalars have also been considered.
References \cite{Venugopal:1998fq, Holstein:2001bt} apply an Omnes function on top of the one-loop results to include the effects of $p$-wave pion scattering. 
Another approach combines ChPT with dispersion theory allowing for a controlled inclusion of resonance physics \cite{Stollenwerk:2011zz}. Due to the inclusion of pion-pion rescattering in the final state both the $\eta$ and the $\eta'$ spectrum can be described well. Reference \cite{Kubis:2015sga} augments this analysis of the $\eta\to\pi^+\pi^-\gamma$ decay by the $a_2$ tensor meson. 
Finally, Ref.~\cite{Dai:2017tew} performed an amplitude analysis of the decay $\eta'\to\pi^+\pi^-\gamma$ based on the latest BESIII data \cite{Ablikim:2017fll}, taking into account $\rho$-$\omega$ mixing.

\section{$\eta^{(')}\to\pi^+\pi^-l^+l^-$}\label{sec:etapipill}

In the following, we investigate the decays involving a virtual photon $\eta^{(')}\to\pi^+\pi^-\gamma^\ast$, which are connected to the decays $\eta^{(')}\to\pi^+\pi^-l^+l^-$, with a lepton pair $l=e,\ \mu$.
The matrix element for the decay $\eta^{(')}\to\pi^+\pi^-\gamma^*$ is given by
\begin{align}
	\mathcal{M}=-iF_P\epsilon_{\mu\nu\alpha\beta}\epsilon^\mu p^\nu_+ p^\alpha_- q^\beta,
	\label{ampEtaPiPiGamma}
\end{align}
where $q^\mu$ and $\epsilon^\mu$ denote the momentum and polarization vector of the photon, respectively, and where $p^\mu_+$, $p^\mu_-$ are the momenta of the pions.
The decay $\eta^{(')}\to\pi^+\pi^-l^+l^-$  proceeds via a two-step mechanism \cite{Picciotto:1993aa,Borasoy:2007dw}. The first decay is $\eta^{(')}\to\pi^+\pi^-\gamma^*$ which is followed by $\gamma^*\to l^+l^-$. We can obtain the invariant amplitude for $\eta^{(')}\to\pi^+\pi^-l^+l^-$ from a modification of the one in Eq.~(\ref{ampEtaPiPiGamma}). The photon is now off shell and we replace its polarization vector $\epsilon^\mu$ by $(e/q^2)\bar{u}(k^-)\gamma^\mu v(k^+)$, where $k^\pm$ are the lepton momenta.
After this modification, the invariant amplitude reads
\begin{align}
	\mathcal{M}=-iF_P\epsilon_{\mu\nu\alpha\beta} p^\nu_+ p^\alpha_- q^\beta \left[\frac{e}{q^2}\bar{u}(k^-)\gamma^\mu v(k^+)\right].
\end{align}
The form factors $F_P$ have been calculated in Sec.~\ref{sec:CalcM}.
We can then calculate the differential decay rates of $\eta^{(')}\to\pi^+\pi^-l^+l^-$ in terms of the normalized invariant mass of the pion pair $x=(p^++p^-)^2/M^2_P\equiv s_{\pi\pi}/M^2_P$ and the normalized invariant mass of the lepton pair $y=(k^++k^-)^2/M^2_P\equiv q^2/M^2_P$, where $P=\eta,\ \eta'$. The differential decay width is given by \cite{Picciotto:1993aa}
\begin{align}
	\frac{d^2\Gamma}{dxdy}=\frac{e^2M^7_P}{18(4\pi)^5}\frac{\lambda^{3/2}(1,x,y)\lambda^{1/2}(y,\nu^2,\nu^2)\lambda^{3/2}(x,\mu^2,\mu^2)}{x^2y^2}\left(\frac{1}{4}+\frac{\nu^2}{2y}\right)\left|F_P\right|^2,
\end{align}
where $\lambda(x,y,z)=x^2+y^2+z^2-2xy-2xz-2yz$ is the K\"all\'{e}n function, $\mu=M_\pi/M_P$, and $\nu=m_l/M_P$.
The spectrum with respect to $x$ is obtained by integrating over $y$ 
\begin{align}
	\frac{d\Gamma}{dx}=\int^{1-2 \sqrt{x}+x}_{4m^2_l/M^2_P} dy \frac{d^2\Gamma}{dxdy},
\end{align}
whereas the integration over $x$ leads to the spectrum with respect to $y$
\begin{align}
	\frac{d\Gamma}{dy}=\int^{1-2 \sqrt{y}+y}_{4M^2_\pi/M^2_P} dx \frac{d^2\Gamma}{dxdy}.
\end{align}
The full decay width of $\eta^{(')}\to\pi^+\pi^-l^+l^-$ is given by
\begin{align}
	\Gamma_{P\to\pi^+\pi^-l^+l^-}=\int^{1-2 \sqrt{4m^2_l/M^2_P}+4m^2_l/M^2_P}_{4M^2_\pi/M^2_P}dx \int^{1-2 \sqrt{x}+x}_{4m^2_l/M^2_P} dy \frac{d^2\Gamma}{dxdy}.
\end{align}

\subsection{Numerical analysis}

While at LO the numerical evaluation of the results can be performed directly, at NLO we need to fix four constants $c_3$, $c_4$, $c_{15}$, and $c_{14}$. For the parameters $c_3$, $c_4$, $c_{15}$ we employ the values determined from the decays to real photons $\eta^{(')}\to\pi^+\pi^-\gamma$ at NLO in Table \ref{tab:FitParametersAtNLO}. The parameter $c_{14}$ is multiplied by the photon virtuality $q^2$ and needs to be fixed to the decays $\eta^{(')}\to\pi^+\pi^-l^+l^-$ involving a virtual photon. The available data for these decays are the decay widths for $\eta^{(')}\to\pi^+\pi^-e^+e^-$ \cite{Zyla:2020zbs} and $\eta'\to\pi^+\pi^-\mu^+\mu^-$ \cite{Ablikim:2020svz}, whereas for the decay width of $\eta\to\pi^+\pi^-\mu^+\mu^-$ only an upper limit exists \cite{Zyla:2020zbs}. The spectra of these decays have not been measured. Since we are not able to describe the full $\eta'\to\pi^+\pi^-\gamma$ spectrum due to the importance of resonant contributions, we expect that the description of the $\eta'\to\pi^+\pi^-e^+e^-$ decay is not appropriate in our framework. However, in the decay $\eta'\to\pi^+\pi^-\mu^+\mu^-$, both a pion pair and a muon pair has to be created, such that their invariant masses do not reach values where the contributions of vector mesons start dominating. Therefore, we can use the decay widths of $\eta\to\pi^+\pi^-e^+e^-$ and $\eta'\to\pi^+\pi^-\mu^+\mu^-$ to determine $c_{14}$. The LEC $c_4$ is set to the three different values determined in Table \ref{tab:FitParametersAtNLO}, corresponding to the different fit ranges for the $\eta'\to\pi^+\pi^-\gamma$ spectrum. We then fix $c_{14}$ through a fit to the experimental data $\Gamma_{\eta\to\pi^+\pi^-e^+e^-}=(351 \pm20)\ \text{meV}$ \cite{Zyla:2020zbs} and $\Gamma_{\eta'\to\pi^+\pi^-\mu^+\mu^-}=(3.70 \pm 0.98)\ \text{eV}$ \cite{Ablikim:2020svz}. The results for $c_{14}$ are displayed in Table \ref{tab:c14fit}.
\begin{table}[htbp]
	\centering
	\begin{tabular}{c c r@{$\,\pm\,$}l}\myTR
	Fit	& $c_4$  & 	\multicolumn{2}{c}{$c_{14}\ [\text{GeV}^{-2}]$}	\\\myMR
NLO I & -0.86 & -3.92 & 3.19 \\
NLO II & -1.24 & -7.45 & 3.11 \\
NLO III & -1.89 & -13.24 & 3.00 \\
		\myBR
	\end{tabular}
	\caption{Results for the fit parameters.}
	\label{tab:c14fit}
\end{table}
As the absolute value of $c_4$ increases, the absolute value of $c_{14}$ gets larger as well. A naive VMD estimate for $c_{14}$ is given by $c_{14}=-2.53\ \text{GeV}^{-2}$ \cite{Hacker:2008}
, which is roughly of the same order of magnitude as our values.

\begin{figure}[htbp]
	\centering
	\includegraphics[width=1.0\textwidth]{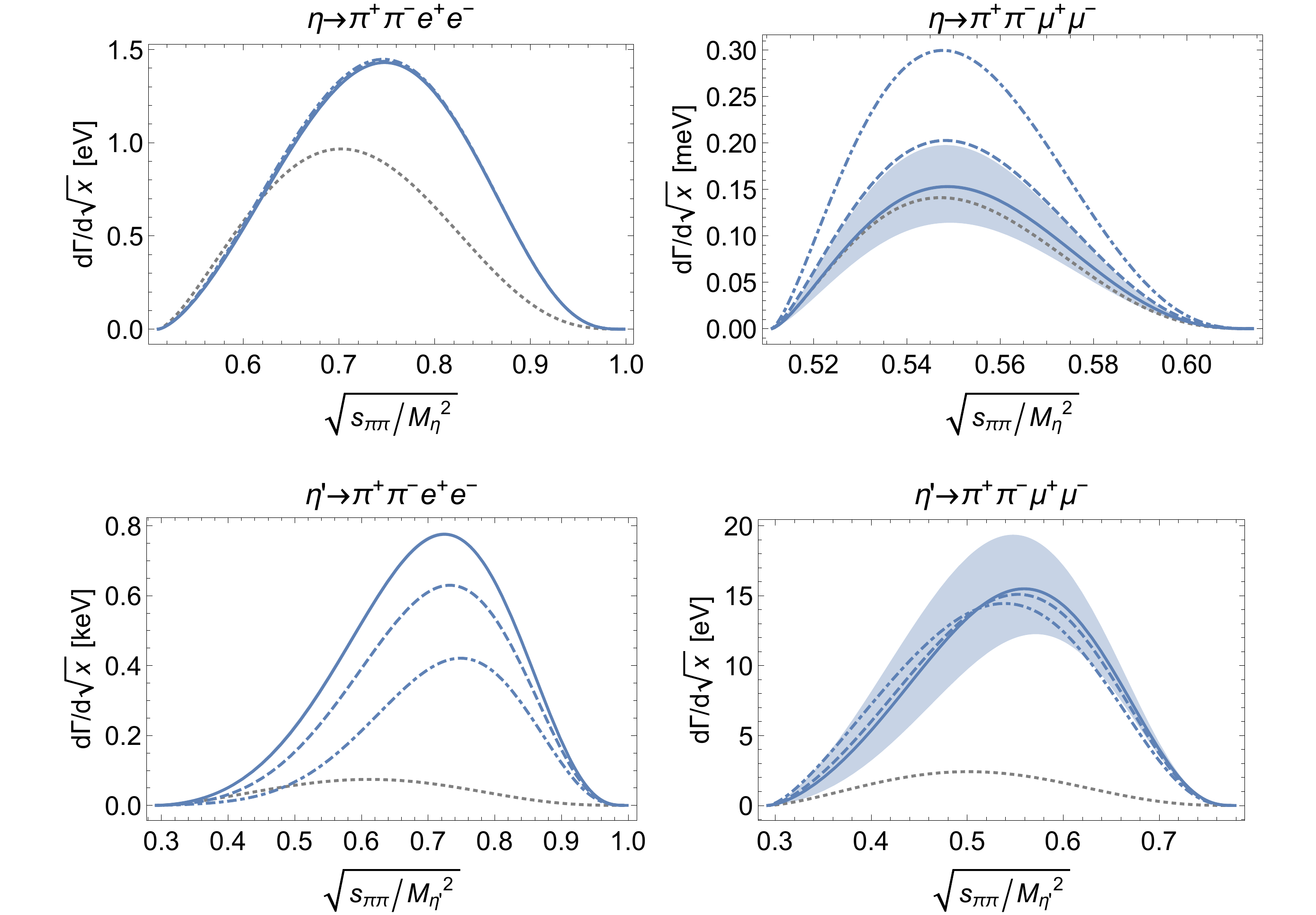}
	\caption{Invariant-mass spectra of the $\pi^+\pi^-$ system at LO (dotted, gray), NLO I (solid, blue), NLO II (dashed, blue), and NLO III (dash-dotted, blue). The bands correspond to the fit error of $c_{14}$ for NLO I.}
	\label{fig:nlos}
\end{figure}

\begin{figure}[htbp]
	\centering
	\includegraphics[width=1.0\textwidth]{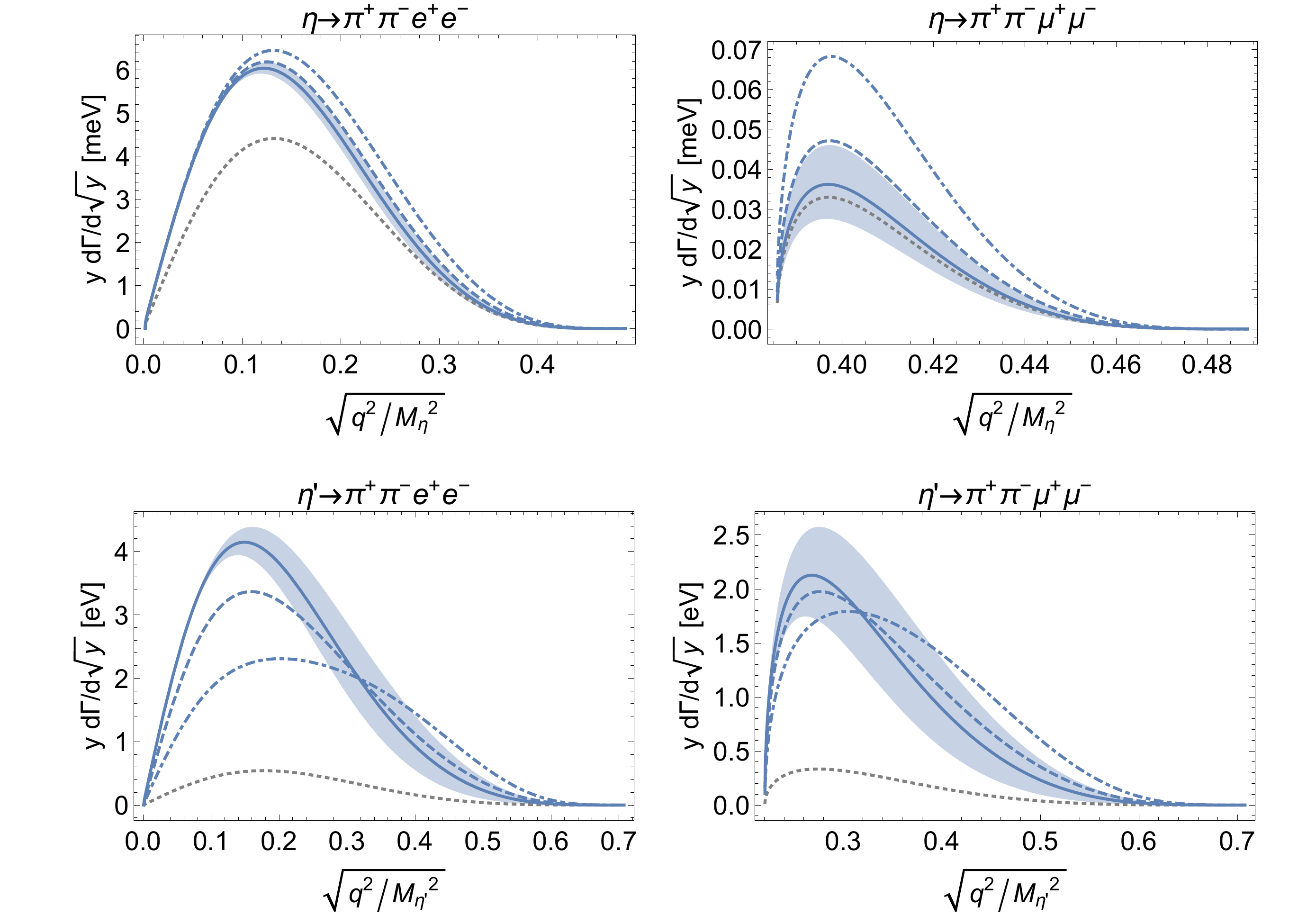}
	\caption{Invariant-mass spectra of the $l^+l^-$ system at LO (dotted, gray), NLO I (solid, blue), NLO II (dashed, blue), and NLO III (dash-dotted, blue). The bands correspond to the fit error of $c_{14}$ for NLO I.}
	\label{fig:nlok}
\end{figure}

In Figs.~\ref{fig:nlos} and \ref{fig:nlok}, we show the predictions for the invariant-mass spectra of the $\pi^+\pi^-$ and $l^+l^-$ systems at NLO for all four decays $\eta^{(')}\to\pi^+\pi^-l^+l^-$, respectively. The spectra are plotted for the three different sets of parameters in Table \ref{tab:c14fit} and are compared to the LO results. To assess the uncertainty in $c_{14}$, for the NLO I fit, we display the error bands resulting from the fit error of $c_{14}$. 


In general, the LO and NLO spectra differ greatly. The NLO corrections tend to produce steeper and larger peaks compared to the LO predictions. For the decays involving an $e^+e^-$ pair, variations of $c_{14}$ have only a minor influence, because the error bands coincide with the line thickness in Fig.~\ref{fig:nlos}. A larger effect can be seen in the invariant-mass spectra of the $l^+l^-$ system in Fig.~\ref{fig:nlok}. The error bands are much larger for the decays to $\mu^+\mu^-$. Due to the larger invariant mass of the muon pair, the photon virtuality is increased and the decays are more sensitive to $c_{14}$. Since the fits are performed to the decay width of $\eta'\to\pi^+\pi^-\mu^+\mu^-$, the three NLO curves are close together, whereas in $\eta\to\pi^+\pi^-\mu^+\mu^-$ the effect of the different $c_{14}$ values can be seen and in $\eta'\to\pi^+\pi^-e^+e^-$ the influence of $c_4$ can be observed.

At NNLO, in addition to the parameters determined from $\eta^{(')}\to\pi^+\pi^-\gamma$, more unknown LECs appear, multiplying possible structures in the form factors like ${(q^2)}^2$ or $q^2s_{\pi\pi}$. 
Therefore, we do not numerically evaluate the full NNLO expressions. At this order, the loops start contributing. For completeness, in order to provide an estimate of the size of the loop corrections, we evaluate the scenario where we just add the loops to the LO expressions. The corresponding spectra are shown in Figs.~\ref{fig:los} and \ref{fig:lok} in Appendix \ref{app:plots}. We observe rather large effects of the loops on the spectra, comparable in size to the NLO corrections. 

Finally, we integrate the spectra and obtain predictions for the full decay widths of $\eta'\to\pi^+\pi^-l^+l^-$. The results are displayed in Table \ref{tab:Widthpipill}. Since this is only a first study of the decays $\eta^{(')}\to\pi^+\pi^-l^+l^-$ to obtain a rough estimate of the higher-order corrections, we do not provide errors for the results of the decay widths.   
\begin{table}[htbp]
	\centering
	\begin{tabular}{ccccc}\myTR
		& $\Gamma_{\eta\to\pi^+\pi^-e^+e^-}$ & $\Gamma_{\eta'\to\pi^+\pi^-e^+e^-}$ & $\Gamma_{\eta\to\pi^+\pi^-\mu^+\mu^-}$ & $\Gamma_{\eta'\to\pi^+\pi^-\mu^+\mu^-}$ \\ 
		&  $[10^{-10}\ \text{GeV}]$&  $[10^{-7}\ \text{GeV}]$&  $[10^{-15}\ \text{GeV}]$&  $[10^{-9}\ \text{GeV}]$		\\\myMR
		\text{LO} & 2.34 & 0.26 & 7.20 & 0.59 \\
		\text{NLO I} & 3.48 & 2.38 & 7.91 & 3.72 \\
		\text{NLO II} & 3.50 & 1.88 & 10.44 & 3.71 \\
		\text{NLO III} & 3.53 & 1.18 & 15.38 & 3.69 \\
		\text{LO+Loops} & 1.81 & 1.13 & 5.16 & 2.50 \\ \myMR
		\text{Experiment \cite{Zyla:2020zbs, Ablikim:2020svz}} & $3.5\pm 0.2$ & $4.5\pm2.4$ & $<4.7\cdot 10^5$ & $3.7\pm1.0$\\ \myMR
		\text{VMD \cite{Picciotto:1993aa}} & 3.8 & - & - & - \\
		\text{\cite{Faessler:1999de}} & 4.72 & 3.56 & 15.72 & 3.96 \\
		\text{CC \cite{Borasoy:2007dw}} & $3.89^{+0.10}_{-0.13}$ & $4.31^{+0.38}_{-0.64}$ & $9.8^{+5.8}_{-3.5}$ & $3.2^{+2.0}_{-1.6}$ \\
		\text{Hidden gauge \cite{Petri:2010ea}} & $4.11\pm0.27$ & $4.3\pm0.46$ & $11.33\pm0.67$ & $4.36\pm0.63$ \\
		\text{Modif. VMD \cite{Petri:2010ea}} & $3.96\pm0.22$ & $4.49\pm0.33$ & $11.32\pm0.54$ & $4.77\pm0.54$\\
		\myBR
	\end{tabular}
	\caption{Results for the decay widths of $\eta^{(')}\to\pi^+\pi^-l^+l^-$.}
	\label{tab:Widthpipill}
\end{table}
The widths of $\eta\to\pi^+\pi^-e^+e^-$ and $\eta'\to\pi^+\pi^-\mu^+\mu^-$ are very well described by the NLO I-III fits. 
In general, the LO values for all decays are quite small and the NLO corrections provide increased results. 
For both $\eta$ decay widths the loop corrections lead to a decrease of about 25\% compared to the LO values, whereas the loops add large positive contributions to the LO results for the $\eta'$ decay widths.
The LO value for $\Gamma_{\eta'\to\pi^+\pi^-e^+e^-}$ is very small. The NLO results depend quite strongly on the different values determined for $c_4$ and are only up to 50\% of the experimental value. This is related to the importance of vector mesons, which we have not taken into account explicitly. Furthermore, the full NNLO contributions might further improve our result. 
For $\eta\to\pi^+\pi^-\mu^+\mu^-$, the experimental limit is five orders of magnitude larger than our determinations.

\subsubsection{Comparison with other works}

In Table \ref{tab:Widthpipill}, we compare our results for the decay widths with other theoretical predictions. In Ref.~\cite{Picciotto:1993aa}, the decay $\eta\to\pi^+\pi^-e^+e^-$ has been studied in a chiral model that incorporates vector mesons explicitly. Reference \cite{Faessler:1999de} calculated various decays of light unflavored mesons using a meson-exchange model based on VMD. A chiral unitary approach that combines ChPT with a coupled-channel Bethe-Salpeter equation has been applied to the decays $\eta^{(')}\to\pi^+\pi^-l^+l^-$ in Ref.~\cite{Borasoy:2007dw}. Reference \cite{Petri:2010ea} investigates the decays within the hidden gauge and a modified VMD model. The results of Refs.~\cite{Borasoy:2007dw,Petri:2010ea} agree within their errors which are quite large in some cases, and the agreement is better for the decays involving $e^+e^-$ than for those with $\mu^+\mu^-$. The results of Ref.~\cite{Faessler:1999de} show larger deviations. Our NLO results for $\Gamma_{\eta\to\pi^+\pi^-e^+e^-}$ are  smaller than the other theoretical values which are larger than the experimental value.  
The other theoretical predictions agree within errors with the experimental value for $\Gamma_{\eta'\to\pi^+\pi^-e^+e^-}$, but they are slightly smaller and Ref.~\cite{Faessler:1999de} shows the greatest deviation. All theory values for $\Gamma_{\eta\to\pi^+\pi^-\mu^+\mu^-}$ are below the experimental limits, while the predictions for $\Gamma_{\eta'\to\pi^+\pi^-\mu^+\mu^-}$ are larger than the experimental value in some cases, but all of them agree within errors.
In general, our NLO results for $\Gamma_{\eta'\to\pi^+\pi^-e^+e^-}$ are substantially lower than the other theoretical predictions. This can be explained by the fact that we, as opposed to the other works, have not taken the explicit contributions of vector mesons into account.

References \cite{Picciotto:1993aa,Borasoy:2007dw,Petri:2010ea} also provide plots of their predicted spectra. The invariant-mass spectra of the $\pi^+\pi^-$ and $e^+e^-$ systems in $\eta\to\pi^+\pi^-e^+e^-$ agree with each other and with our NLO results for the spectra.  
For the spectra of $\eta\to\pi^+\pi^-\mu^+\mu^-$ with respect to $\sqrt{s_{\pi\pi}}$ and $\sqrt{q^2}$, we find qualitative agreement of our NLO results with Refs.~\cite{Borasoy:2007dw,Petri:2010ea}, with the difference that  
our peaks are a little bit higher than those of the other works. Our NLO $\pi^+\pi^-$ invariant-mass spectrum of $\eta'\to\pi^+\pi^-e^+e^-$ is much broader and lower than those in Refs.~\cite{Borasoy:2007dw,Petri:2010ea}, which exhibit a steep peak around $750$ MeV. Less pronounced is the behavior in the $e^+e^-$ invariant-mass spectrum, but also there our peak is broader and lower. Here, the influence of the explicit vector mesons which are included in Refs.~\cite{Borasoy:2007dw,Petri:2010ea} can be clearly seen. With regard to the spectra for $\eta'\to\pi^+\pi^-\mu^+\mu^-$, our results agree quite well with Ref.~\cite{Borasoy:2007dw},
except that our peak in the invariant-mass spectrum of the $\mu^+\mu^-$ system is broader than in Ref.~\cite{Borasoy:2007dw}.

In order to test the different approaches to the decays $\eta^{(')}\to\pi^+\pi^-l^+l^-$, more experimental data on the decays is highly desirable. Experimental data on the differential decay spectra of any of the decays $\eta^{(')}\to\pi^+\pi^-l^+l^-$ or the decay width of $\eta\to\pi^+\pi^-\mu^+\mu^-$ would allow for an improved determination of the parameter $c_{14}$ and might even facilitate the determination of LECs at NNLO.

\section{Summary and outlook}\label{sec:Conclusions}

We have investigated the decays $\eta^{(')}\to\pi^+\pi^-\gamma^{(\ast)}$ at the one-loop level up to and including NNLO in L$N_c$ChPT. Besides the loop corrections, all contact terms up to and including NNLO have been taken into account. To this end, possible structures from the $\mathcal{O}(p^8)$ Lagrangian, which has not been constructed yet, have been introduced phenomenologically, together with free parameters. In addition, the $\eta-\eta'$ mixing has been consistently included. We have numerically evaluated the decays successively at LO, NLO, and NNLO. For $\eta^{(')}\to\pi^+\pi^-\gamma$, the LECs from the odd-intrinsic-parity sector were determined through fits to the decay width and the full decay spectrum of the $\eta$ and to parts of the $\eta'$ decay spectrum, since we are not able to adequately describe the full $\eta'$ spectrum. 
In general, the results for the spectra gradually improve from LO, which is far off, to NLO and NNLO. In the case of the $\eta$, the experimental data are well described at NNLO, mainly due to the higher-order contact terms, while the loop corrections have only a very small influence. For the $\eta'$ decay, the loops are more important and the $s^2_{\pi\pi}$ term is only relevant at high values of the $\pi^+\pi^-$ invariant mass, leading to a good description of the $\eta'$ spectrum up to $\sqrt{s_{\pi\pi}}=0.7\ \text{GeV}$. Here, our approach reaches its limit, since resonant contributions of vector mesons become important.
Finally, we have considered the decays $\eta^{(')}\to\pi^+\pi^-l^+l^-$, $l=e,\ \mu$. At NLO, the LEC $c_{14}$, which accompanies the photon virtuality, could be fixed to the decay widths of $\eta\to\pi^+\pi^-e^+e^-$ and $\eta'\to\pi^+\pi^-\mu^+\mu^-$. We have then evaluated the decay spectra of all four decays with respect to the invariant masses of the $\pi^+\pi^-$ and $l^+l^-$ systems at NLO. The NLO corrections modify the spectra substantially in comparison with the LO results. Unfortunately no experimental data for the spectra are available. We have compared our results with other theoretical determinations and find agreement in some cases. Discrepancies arise when vector-meson degrees of freedom play a role, which have been taken into account in the other works. 
At NNLO, due to the appearance of additional unknown LECs, we only evaluated the spectra for the scenario where the loop corrections were added to the LO results. We have found that the loop contributions are of the same order of magnitude as the NLO corrections. 
To further test the various theoretical approaches, more experimental information on the differential spectra of any of the four decays or on the decay widths of $\eta\to\pi^+\pi^-\mu^+\mu^-$ would be very helpful, since it would allow for a better determination of the LECs at NLO and maybe even at NNLO. 

Our results show the limitations of a perturbative chiral and large $N_c$ expansion, especially in the case of the $\eta'\to\pi^+\pi^-\gamma$ spectrum. While the extension to higher orders might further improve the description of the data, the number of unknown LECs increases, thus making the gain in physical insight questionable. However, the inclusion of vector mesons as explicit degrees of freedom might extend the range of applicability of the effective theory.

\begin{acknowledgements}
Supported by the Deutsche Forschungsgemeinschaft DFG through the Collaborative Research Center “The Low-Energy Frontier of the Standard Model” (SFB 1044). 
\end{acknowledgements}

\begin{appendix}

\section{Additional expressions}\label{app:Expressions}

The loop contributions to the form factors of the decays $\eta^{(')}\to\pi^+\pi^-\gamma^*$ given by the loop diagrams in Fig.~\ref{fig:phigallp} read
\begin{align}
F_\eta
&=\frac{1}{768 \sqrt{3} \pi ^4 F_{\pi
	}^5}\left(2 \cos (\theta^{[0]} ) \left[3 \left(q^2-4 M_K^2\right) B_0\left(q^2,M_K^2,M_K^2\right)\right.\right.\no\\
&\quad+2
\left(s_{\pi\pi}-4 M_K^2\right) B_0\left(s_{\pi\pi},M_K^2,M_K^2\right)+\left(s_{\pi\pi}-4 M_{\pi }^2\right)
B_0\left(s_{\pi\pi},M_{\pi }^2,M_{\pi }^2\right)\no\\
&\quad\left.+2 A_0\left(M_K^2\right)+22 A_0\left(M_{\pi
}^2\right)+2 \left(-10 M_K^2-2 M_{\pi }^2+q^2+s_{\pi\pi}\right)\right]\no\\
&\quad-\sqrt{2} \sin (\theta^{[0]} )
\left\{\left(s_{\pi\pi}-4 M_K^2\right) B_0\left(s_{\pi\pi},M_K^2,M_K^2\right)+2 \left[\left(s_{\pi\pi}-4 M_{\pi
}^2\right) \right.\right.\no\\
&\quad\left.\left.\left.\times B_0\left(s_{\pi\pi},M_{\pi }^2,M_{\pi }^2\right)-2 M_K^2-4 M_{\pi }^2+s_{\pi\pi}\right]+22
A_0\left(M_K^2\right)+44 A_0\left(M_{\pi }^2\right)\right\}\right)
\end{align}
and
\begin{align}
F_{\eta'}
&=\frac{1}{768 \sqrt{3} \pi ^4 F_{\pi
	}^5}\left(2 \sin (\theta^{[0]} ) \left\{3 \left(q^2-4 M_K^2\right) B_0\left(q^2,M_K^2,M_K^2\right)\right.\right.\no\\
&\quad+2
\left(s_{\pi\pi}-4 M_K^2\right) B_0\left(s_{\pi\pi},M_K^2,M_K^2\right)+\left(s_{\pi\pi}-4 M_{\pi }^2\right)
B_0\left(s_{\pi\pi},M_{\pi }^2,M_{\pi }^2\right)\no\\
&\quad\left.+2 A_0\left(M_K^2\right)+22 A_0\left(M_{\pi
}^2\right)+2 \left[-2 \left(5 M_K^2+M_{\pi }^2\right)+q^2+s_{\pi\pi}\right]\right\}\no\\
&\quad+\sqrt{2} \cos
(\theta^{[0]} ) \left\{\left(s_{\pi\pi}-4 M_K^2\right) B_0\left(s_{\pi\pi},M_K^2,M_K^2\right)+2 \left[\left(s_{\pi\pi}-4
M_{\pi }^2\right) \right.\right.\no\\
&\quad\left.\left.\left.\times B_0\left(s_{\pi\pi},M_{\pi }^2,M_{\pi }^2\right)-2 M_K^2-4 M_{\pi }^2+s_{\pi\pi}\right]+22
A_0\left(M_K^2\right)+44 A_0\left(M_{\pi }^2\right)\right\}\right).
\end{align}
The explicit expressions for the loop integrals read
\begin{align}
A_0(m^2)&=(-16\pi^2)\left[2m^2\lambda+\frac{m^2}{8\pi^2}\text{ln}\left(\frac{m}{\mu}\right)\right],\\
B_0(p^2,m^2_1,m^2_2)&=(-16\pi^2)\left\{2\lambda+\frac{\text{ln}\left(\frac{m_1}{\mu}\right)}{8\pi^2}+\frac{1}{16\pi^2}\right.\no\\
&\times\left.\left[-1+\frac{p^2-m^2_1+m^2_2}{p^2}\text{ln}\left(\frac{m_2}{m_1}\right)+\frac{2m_1m_2}{p^2}F(\Omega)\right]\right\},
\end{align}
where
\begin{align}
\lambda&=\frac{1}{16\pi^2}\left\{\frac{1}{n-4}-\frac{1}{2}[\text{ln}(4\pi)+\Gamma'(1)+1]\right\},\\
\Omega&=\frac{p^2-m^2_1-m^2_2}{2m_1m_2}
\end{align}
and
\begin{equation}
F(\Omega)=\begin{cases} \sqrt{\Omega^2-1}\;\text{ln}\left(-\Omega-\sqrt{\Omega^2-1}\right) & \text{for}\ \ \Omega\leq-1,\\
\sqrt{1-\Omega^2}\;\text{arccos}\left(-\Omega\right)               & \text{for}\ \ -1\leq\Omega\leq1,\\
\sqrt{\Omega^2-1}\;\text{ln}\left(\Omega+\sqrt{\Omega^2-1}\right)-i\pi\sqrt{\Omega^2-1} & \text{for}\ \ 1\leq\Omega.
\end{cases}
\end{equation}
We evaluate the loop integrals at the renormalization scale $\mu=1\ \text{GeV}$.

\section{Fit parameters}\label{app:Par}

\begin{table}[htbp]
	\centering
	\begin{tabular}{c r@{$\,\pm\,$}l r@{$\,\pm\,$}l r@{$\,\pm\,$}l r@{$\,\pm\,$}l c}\myTR
 & \multicolumn{2}{c}{$A_\eta$ $[10^{10}]$} & \multicolumn{2}{c}{$b_\eta$} & \multicolumn{2}{c}{$c_\eta\ [\text{GeV}^{-2}]$} & \multicolumn{2}{c}{$d_\eta\ [\text{GeV}^{-4}]$} & MSE \\ \myMR
\text{Full} & $1.29$&$0.05$ & $0.09$&$0.17$ & $-4.60$&$2.03$ & $34.35$&$6.05$ & 1.10 \\
\text{Without loops} & $1.45$&$0.06$ & $-0.01$&$0.16$ & $-3.30$&$1.92$ & $31.49$&$5.72$ & 1.11 \\
$d_{\eta }=0$ & $1.28$&$0.05$ & $-2.03$&$0.05$ & $-8.41$&$0.30$ & $0.$&$0.$ & 1.88 \\
\text{Without loops $\land $ }$d_{\eta }=0$ & $1.43$&$0.06$ & $-0.84$&$0.04$ & $7.24$&$0.29$ & $0.$&$0.$ & 1.94 \\
		\myBR
	\end{tabular}
	\caption{Fit parameters for the $\eta$ spectrum at NNLO determined in Sec.~\ref{sec:etapipigammaNA}.}
	\label{tab:NNLOeta}
\end{table}

\begin{table}[htbp]
	\centering
	\begin{tabular}{c r@{$\,\pm\,$}l r@{$\,\pm\,$}l r@{$\,\pm\,$}l r@{$\,\pm\,$}l c}\myTR
		& \multicolumn{2}{c}{$A_{\eta'}$ $[10^{10}]$} & \multicolumn{2}{c}{$b_{\eta'}$} & \multicolumn{2}{c}{$c_{\eta'}\ [\text{GeV}^{-2}]$} & \multicolumn{2}{c}{$d_{\eta'}\ [\text{GeV}^{-4}]$} & MSE \\\myMR
\text{Full I} & $-0.19$&$0.00$ & $4.57$&$0.18$ & $-0.69$&$0.01$ & $-1.11$&$0.09$ & 0.8 \\
\text{Full II} & $-8.49$&$0.03$ & $1.55$&$0.04$ & $-1.02$&$0.00$ & $-0.88$&$0.02$ & 0.77 \\
\text{Full III} & $-8.39$&$0.02$ & $1.78$&$0.01$ & $-1.01$&$0.00$ & $-0.99$&$0.01$ & 1.59 \\
$d_{\eta '}$\text{=0 I} & $-8.05$&$0.08$ & $-0.96$&$0.00$ & $-0.48$&$0.01$ & $0.$&$0.$ & 0.83 \\
$d_{\eta '}$\text{=0 II} & $-8.68$&$0.02$ & $-0.95$&$0.00$ & $-0.53$&$0.00$ & $0.$&$0.$ & 3.45 \\
$d_{\eta '}$\text{=0 III} & $-8.78$&$0.03$ & $-0.92$&$0.00$ & $-0.68$&$0.00$ & $0.$&$0.$ & 73.16 \\
		\myBR
	\end{tabular}
	\caption{Fit parameters for the $\eta'$ spectrum at NNLO including loops determined in Sec.~\ref{sec:etapipigammaNA}.}
	\label{tab:NNLOetapl}
\end{table}

\begin{table}[htbp]
	\centering
	\begin{tabular}{c r@{$\,\pm\,$}l r@{$\,\pm\,$}l r@{$\,\pm\,$}l c}\myTR
		& \multicolumn{2}{c}{$\tilde{A}_{\eta'}$ $[10^{7}]$} & \multicolumn{2}{c}{$\tilde{c}_{\eta'}\ [\text{GeV}^{-2}]$} & \multicolumn{2}{c}{$\tilde{d}_{\eta'}\ [\text{GeV}^{-4}]$} & MSE \\ \myMR
\text{Without loops I} & $-16.72$&$1.12$ & $-1.71$&$0.23$ & $13.71$&$0.19$ & 0.79 \\
\text{Without loops II} & $-22.82$&$0.91$ & $-2.81$&$0.10$ & $14.66$&$0.09$ & 1.65 \\
\text{Witout loops III} & $-46.73$&$0.83$ & $-4.49$&$0.02$ & $15.04$&$0.06$ & 16.51 \\
\text{Without loops $\land $ }$\tilde{d}_{\eta '}$\text{=0 I} & $-0.97$&$0.08$ & $20.48$&$0.96$ & $0.$&$0.$ & 11.58 \\
\text{Without loops $\land $ }$\tilde{d}_{\eta '}$\text{=0 II} & $-0.03$&$0.01$ & $-156.64$&$30.47$ & $0.$&$0.$ & 46.27 \\
\text{Without loops $\land $ }$\tilde{d}_{\eta '}$\text{=0 III} & $-15.07$&$0.22$ & $-9.84$&$0.05$ & $0.$&$0.$ & 323.05 \\
		\myBR
	\end{tabular}
	\caption{Fit parameters for the $\eta'$ spectrum at NNLO without loops determined in Sec.~\ref{sec:etapipigammaNA}.}
	\label{tab:NNLOetap}
\end{table}

\newpage
\section{Additional plots}\label{app:plots}

\begin{figure}[h!]
	\centering
	\includegraphics[width=1.0\textwidth]{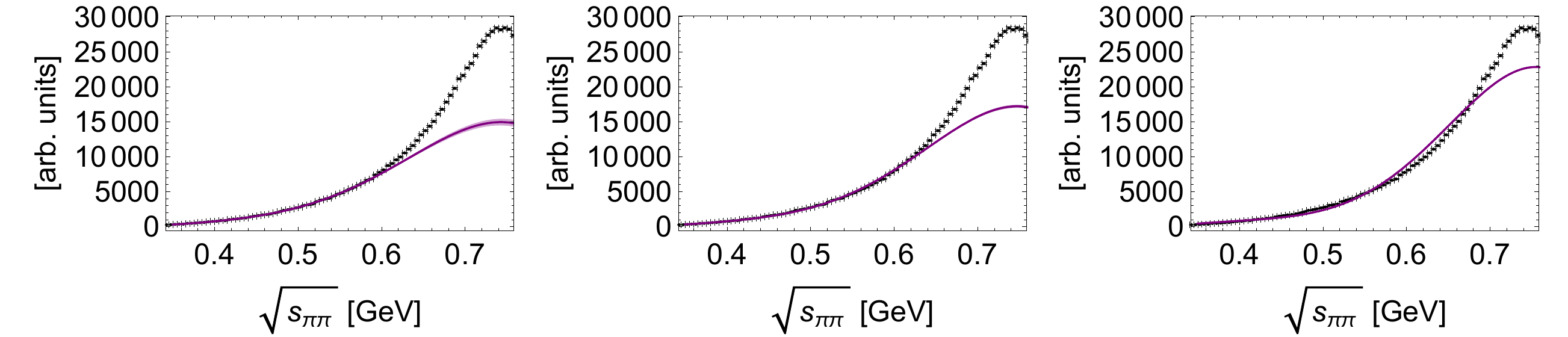}
	\caption{Invariant-mass spectrum of the $\pi^+\pi^-$ system in $\eta'\to\pi^+\pi^-\gamma$ at NNLO with $d_{\eta'}=0$ fitted up to $0.59$ GeV (left), $0.64$ GeV (middle), $0.72$ GeV (right) including the 1$\sigma$ error bands, which partially coincide with the line thickness. The experimental data are taken from Ref.~\cite{Ablikim:2017fll}.}
	\label{fig:b2p}
\end{figure}

\begin{figure}[h!]
	\centering
	\includegraphics[width=1.0\textwidth]{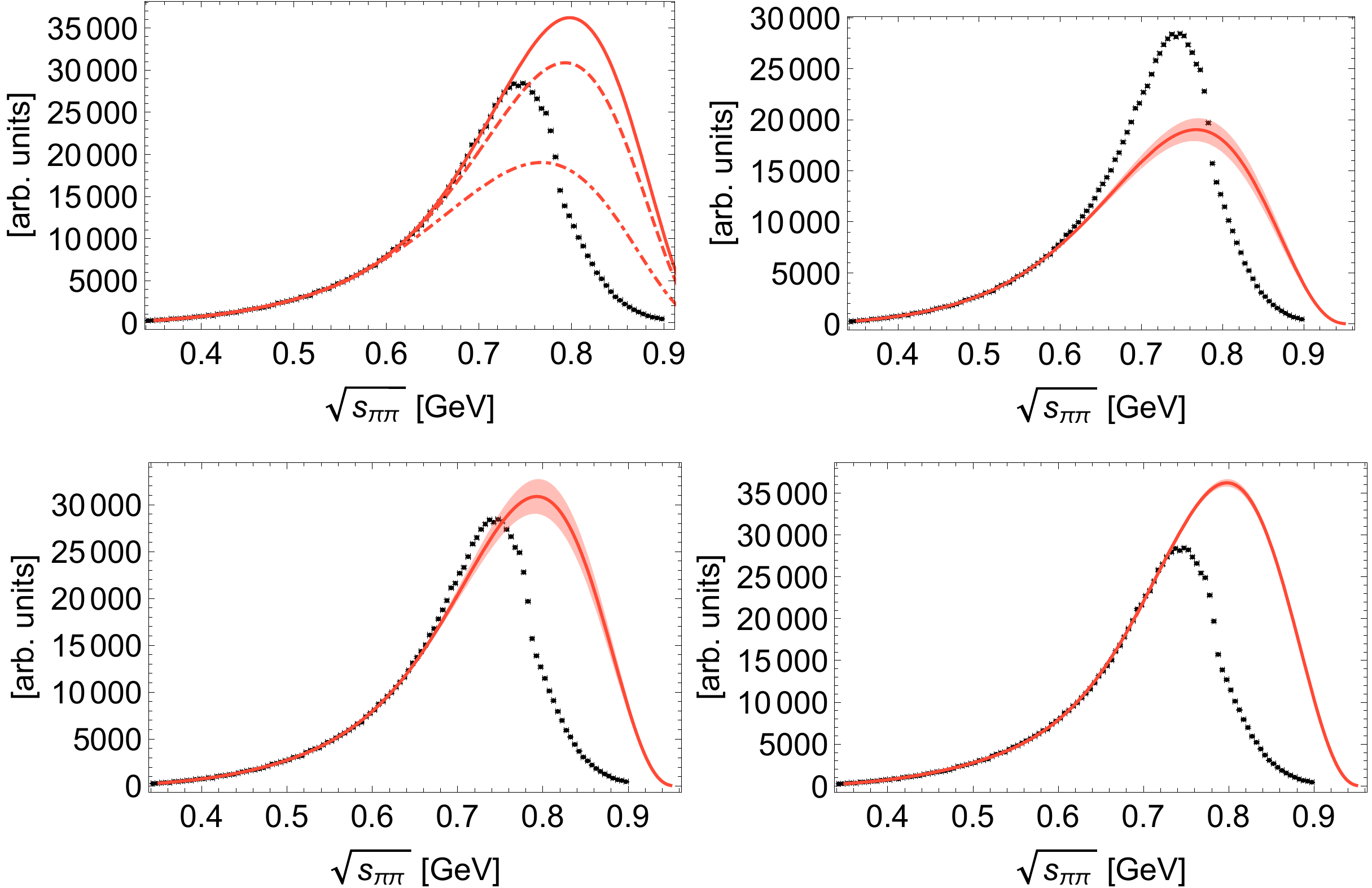}
	\caption{Upper-left plot: Invariant-mass spectrum of the $\pi^+\pi^-$ system in $\eta'\to\pi^+\pi^-\gamma$ at NNLO fitted up to $0.59$ GeV (dash-dotted), $0.64$ GeV (dashed), $0.72$ GeV (solid). Upper-right plot: $1\sigma$ error band for the fit up to $0.59$ GeV. Lower-left plot: $1\sigma$ error band for the fit up to $0.64$ GeV. Lower-right plot: $1\sigma$ error band for the fit up to $0.72$ GeV. The experimental data are taken from Ref.~\cite{Ablikim:2017fll}.}
	\label{fig:b2p1}
\end{figure}

\begin{figure}[h!]
	\centering
	\includegraphics[width=0.85\textwidth]{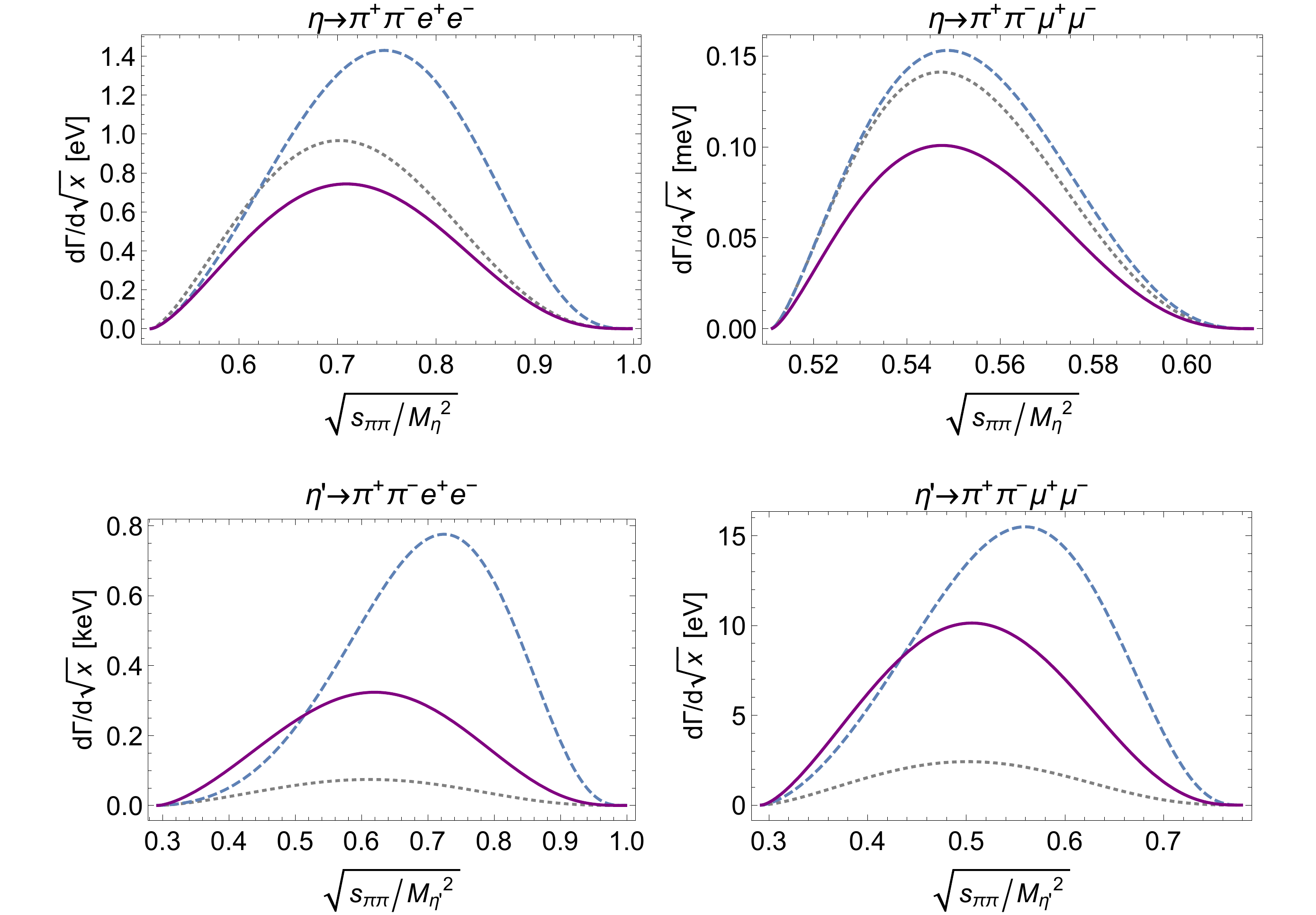}
	\caption{Invariant-mass spectra of the $\pi^+\pi^-$ system at LO (dotted, gray), NLO I (dashed, blue), and LO with loops added (solid, purple).}
	\label{fig:los}
\end{figure}

\begin{figure}[h!]
	\centering
	\includegraphics[width=0.85\textwidth]{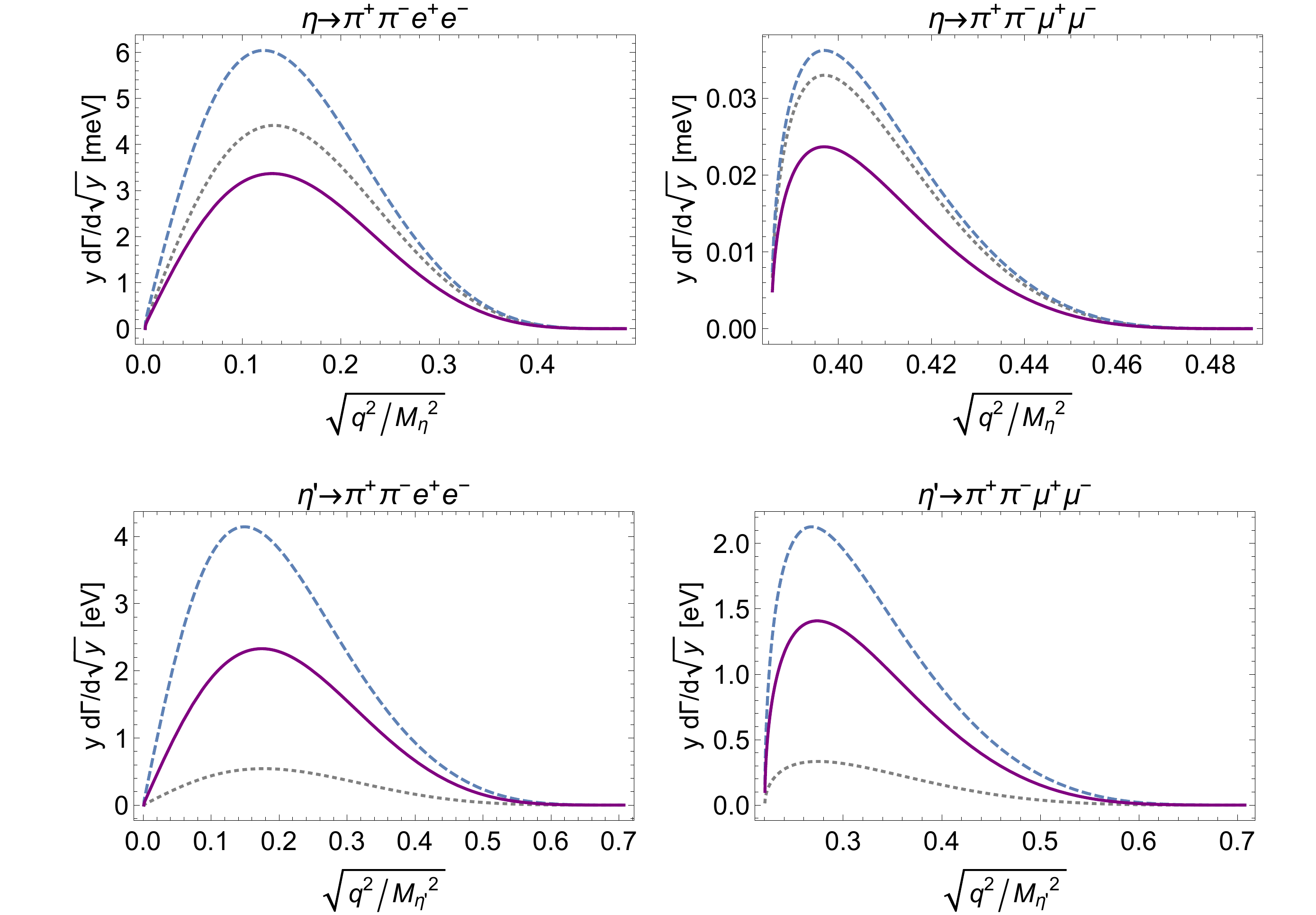}
	\caption{Invariant-mass spectra of the $l^+l^-$ system at LO (dotted, gray), NLO I (dashed, blue), and LO with loops added (solid, purple).}
	\label{fig:lok}
\end{figure}

\clearpage

\end{appendix}

\bibliography{ReferencesPPPgamma.bib}

\begin{thebibliography}{48}%
\makeatletter
\providecommand \@ifxundefined [1]{%
 \@ifx{#1\undefined}
}%
\providecommand \@ifnum [1]{%
 \ifnum #1\expandafter \@firstoftwo
 \else \expandafter \@secondoftwo
 \fi
}%
\providecommand \@ifx [1]{%
 \ifx #1\expandafter \@firstoftwo
 \else \expandafter \@secondoftwo
 \fi
}%
\providecommand \natexlab [1]{#1}%
\providecommand \enquote  [1]{``#1''}%
\providecommand \bibnamefont  [1]{#1}%
\providecommand \bibfnamefont [1]{#1}%
\providecommand \citenamefont [1]{#1}%
\providecommand \href@noop [0]{\@secondoftwo}%
\providecommand \href [0]{\begingroup \@sanitize@url \@href}%
\providecommand \@href[1]{\@@startlink{#1}\@@href}%
\providecommand \@@href[1]{\endgroup#1\@@endlink}%
\providecommand \@sanitize@url [0]{\catcode `\\12\catcode `\$12\catcode
  `\&12\catcode `\#12\catcode `\^12\catcode `\_12\catcode `\%12\relax}%
\providecommand \@@startlink[1]{}%
\providecommand \@@endlink[0]{}%
\providecommand \url  [0]{\begingroup\@sanitize@url \@url }%
\providecommand \@url [1]{\endgroup\@href {#1}{\urlprefix }}%
\providecommand \urlprefix  [0]{URL }%
\providecommand \Eprint [0]{\href }%
\providecommand \doibase [0]{http://dx.doi.org/}%
\providecommand \selectlanguage [0]{\@gobble}%
\providecommand \bibinfo  [0]{\@secondoftwo}%
\providecommand \bibfield  [0]{\@secondoftwo}%
\providecommand \translation [1]{[#1]}%
\providecommand \BibitemOpen [0]{}%
\providecommand \bibitemStop [0]{}%
\providecommand \bibitemNoStop [0]{.\EOS\space}%
\providecommand \EOS [0]{\spacefactor3000\relax}%
\providecommand \BibitemShut  [1]{\csname bibitem#1\endcsname}%
\let\auto@bib@innerbib\@empty
\bibitem [{\citenamefont {Gan}\ \emph {et~al.}(2020)\citenamefont {Gan},
  \citenamefont {Kubis}, \citenamefont {Passemar},\ and\ \citenamefont
  {Tulin}}]{gan2020precision}%
  \BibitemOpen
  \bibfield  {author} {\bibinfo {author} {\bibfnamefont {L.}~\bibnamefont
  {Gan}}, \bibinfo {author} {\bibfnamefont {B.}~\bibnamefont {Kubis}}, \bibinfo
  {author} {\bibfnamefont {E.}~\bibnamefont {Passemar}}, \ and\ \bibinfo
  {author} {\bibfnamefont {S.}~\bibnamefont {Tulin}},\ }\href@noop {} {\enquote
  {\bibinfo {title} {Precision tests of fundamental physics with $\eta$ and
  $\eta^\prime$ mesons},}\ } (\bibinfo {year} {2020}),\ \Eprint
  {http://arxiv.org/abs/2007.00664} {arXiv:2007.00664 [hep-ph]} \BibitemShut
  {NoStop}%
\bibitem [{\citenamefont {Wess}\ and\ \citenamefont
  {Zumino}(1971)}]{Wess:1971yu}%
  \BibitemOpen
  \bibfield  {author} {\bibinfo {author} {\bibfnamefont {J.}~\bibnamefont
  {Wess}}\ and\ \bibinfo {author} {\bibfnamefont {B.}~\bibnamefont {Zumino}},\
  }\href {\doibase https://doi.org/10.1016/0370-2693(71)90582-X} {\bibfield
  {journal} {\bibinfo  {journal} {Physics Letters B}\ }\textbf {\bibinfo
  {volume} {37}},\ \bibinfo {pages} {95} (\bibinfo {year} {1971})}\BibitemShut
  {NoStop}%
\bibitem [{\citenamefont {Witten}(1983)}]{Witten:1983tw}%
  \BibitemOpen
  \bibfield  {author} {\bibinfo {author} {\bibfnamefont {E.}~\bibnamefont
  {Witten}},\ }\href {\doibase https://doi.org/10.1016/0550-3213(83)90063-9}
  {\bibfield  {journal} {\bibinfo  {journal} {Nuclear Physics B}\ }\textbf
  {\bibinfo {volume} {223}},\ \bibinfo {pages} {422} (\bibinfo {year}
  {1983})}\BibitemShut {NoStop}%
\bibitem [{\citenamefont {Stollenwerk}\ \emph {et~al.}(2012)\citenamefont
  {Stollenwerk}, \citenamefont {Hanhart}, \citenamefont {Kupsc}, \citenamefont
  {Mei\ss{}ner},\ and\ \citenamefont {Wirzba}}]{Stollenwerk:2011zz}%
  \BibitemOpen
  \bibfield  {author} {\bibinfo {author} {\bibfnamefont {F.}~\bibnamefont
  {Stollenwerk}}, \bibinfo {author} {\bibfnamefont {C.}~\bibnamefont
  {Hanhart}}, \bibinfo {author} {\bibfnamefont {A.}~\bibnamefont {Kupsc}},
  \bibinfo {author} {\bibfnamefont {U.-G.}\ \bibnamefont {Mei\ss{}ner}}, \ and\
  \bibinfo {author} {\bibfnamefont {A.}~\bibnamefont {Wirzba}},\ }\href
  {\doibase 10.1016/j.physletb.2011.12.008} {\bibfield  {journal} {\bibinfo
  {journal} {Physics Letters B}\ }\textbf {\bibinfo {volume} {707}},\ \bibinfo
  {pages} {184–190} (\bibinfo {year} {2012})}\BibitemShut {NoStop}%
\bibitem [{\citenamefont {Kubis}\ and\ \citenamefont
  {Plenter}(2015)}]{Kubis:2015sga}%
  \BibitemOpen
  \bibfield  {author} {\bibinfo {author} {\bibfnamefont {B.}~\bibnamefont
  {Kubis}}\ and\ \bibinfo {author} {\bibfnamefont {J.}~\bibnamefont
  {Plenter}},\ }\href {http://dx.doi.org/10.1140/epjc/s10052-015-3495-5}
  {\bibfield  {journal} {\bibinfo  {journal} {Eur.~Phys.~J.~C}\ }\textbf
  {\bibinfo {volume} {75}},\ \bibinfo {pages} {283} (\bibinfo {year}
  {2015})}\BibitemShut {NoStop}%
\bibitem [{\citenamefont {Hanhart}\ \emph {et~al.}(2017)\citenamefont
  {Hanhart}, \citenamefont {Holz}, \citenamefont {Kubis}, \citenamefont
  {Kup\'s\'c}, \citenamefont {Wirzba},\ and\ \citenamefont
  {Xiao}}]{Hanhart:2016pcd}%
  \BibitemOpen
  \bibfield  {author} {\bibinfo {author} {\bibfnamefont {C.}~\bibnamefont
  {Hanhart}}, \bibinfo {author} {\bibfnamefont {S.}~\bibnamefont {Holz}},
  \bibinfo {author} {\bibfnamefont {B.}~\bibnamefont {Kubis}}, \bibinfo
  {author} {\bibfnamefont {A.}~\bibnamefont {Kup\'s\'c}}, \bibinfo {author}
  {\bibfnamefont {A.}~\bibnamefont {Wirzba}}, \ and\ \bibinfo {author}
  {\bibfnamefont {C.~W.}\ \bibnamefont {Xiao}},\ }\href {\doibase
  10.1140/epjc/s10052-017-4651-x} {\bibfield  {journal} {\bibinfo  {journal}
  {Eur. Phys. J. C}\ }\textbf {\bibinfo {volume} {77}},\ \bibinfo {pages} {98}
  (\bibinfo {year} {2017})},\ \bibinfo {note} {[Erratum: Eur.Phys.J.C 78, 450
  (2018)]}\BibitemShut {NoStop}%
\bibitem [{\citenamefont {Blum}\ \emph {et~al.}(2016)\citenamefont {Blum},
  \citenamefont {Boyle}, \citenamefont {Izubuchi}, \citenamefont {Jin},
  \citenamefont {J\"uttner}, \citenamefont {Lehner}, \citenamefont {Maltman},
  \citenamefont {Marinkovic}, \citenamefont {Portelli},\ and\ \citenamefont
  {Spraggs}}]{Blum2016}%
  \BibitemOpen
  \bibfield  {author} {\bibinfo {author} {\bibfnamefont {T.}~\bibnamefont
  {Blum}}, \bibinfo {author} {\bibfnamefont {P.~A.}\ \bibnamefont {Boyle}},
  \bibinfo {author} {\bibfnamefont {T.}~\bibnamefont {Izubuchi}}, \bibinfo
  {author} {\bibfnamefont {L.}~\bibnamefont {Jin}}, \bibinfo {author}
  {\bibfnamefont {A.}~\bibnamefont {J\"uttner}}, \bibinfo {author}
  {\bibfnamefont {C.}~\bibnamefont {Lehner}}, \bibinfo {author} {\bibfnamefont
  {K.}~\bibnamefont {Maltman}}, \bibinfo {author} {\bibfnamefont
  {M.}~\bibnamefont {Marinkovic}}, \bibinfo {author} {\bibfnamefont
  {A.}~\bibnamefont {Portelli}}, \ and\ \bibinfo {author} {\bibfnamefont
  {M.}~\bibnamefont {Spraggs}} (\bibinfo {collaboration} {RBC and UKQCD
  Collaborations}),\ }\href {\doibase 10.1103/PhysRevLett.116.232002}
  {\bibfield  {journal} {\bibinfo  {journal} {Phys. Rev. Lett.}\ }\textbf
  {\bibinfo {volume} {116}},\ \bibinfo {pages} {232002} (\bibinfo {year}
  {2016})}\BibitemShut {NoStop}%
\bibitem [{\citenamefont {Jegerlehner}(2017)}]{Jegerlehner:2017gek}%
  \BibitemOpen
  \bibfield  {author} {\bibinfo {author} {\bibfnamefont {F.}~\bibnamefont
  {Jegerlehner}},\ }\href {\doibase 10.1007/978-3-319-63577-4} {\emph {\bibinfo
  {title} {{The Anomalous Magnetic Moment of the Muon}}}},\ Vol.\ \bibinfo
  {volume} {274}\ (\bibinfo  {publisher} {Springer},\ \bibinfo {address}
  {Cham},\ \bibinfo {year} {2017})\BibitemShut {NoStop}%
\bibitem [{\citenamefont {Morte}\ \emph {et~al.}(2017)\citenamefont {Morte},
  \citenamefont {Francis}, \citenamefont {G\"ulpers}, \citenamefont
  {Herdo\'{i}za}, \citenamefont {von Hippel}, \citenamefont {Horch},
  \citenamefont {J\"ager}, \citenamefont {Meyer}, \citenamefont {Nyffeler},\
  and\ \citenamefont {Wittig}}]{Morte_2017}%
  \BibitemOpen
  \bibfield  {author} {\bibinfo {author} {\bibfnamefont {M.~D.}\ \bibnamefont
  {Morte}}, \bibinfo {author} {\bibfnamefont {A.}~\bibnamefont {Francis}},
  \bibinfo {author} {\bibfnamefont {V.}~\bibnamefont {G\"ulpers}}, \bibinfo
  {author} {\bibfnamefont {G.}~\bibnamefont {Herdo\'{i}za}}, \bibinfo {author}
  {\bibfnamefont {G.}~\bibnamefont {von Hippel}}, \bibinfo {author}
  {\bibfnamefont {H.}~\bibnamefont {Horch}}, \bibinfo {author} {\bibfnamefont
  {B.}~\bibnamefont {J\"ager}}, \bibinfo {author} {\bibfnamefont
  {H.}~\bibnamefont {Meyer}}, \bibinfo {author} {\bibfnamefont
  {A.}~\bibnamefont {Nyffeler}}, \ and\ \bibinfo {author} {\bibfnamefont
  {H.}~\bibnamefont {Wittig}},\ }\href
  {http://dx.doi.org/10.1007/JHEP10(2017)020} {\bibfield  {journal} {\bibinfo
  {journal} {J. High Energ. Phys.}\ }\textbf {\bibinfo {volume} {2017}},\
  \bibinfo {pages} {20} (\bibinfo {year} {2017})}\BibitemShut {NoStop}%
\bibitem [{\citenamefont {Herczeg}\ and\ \citenamefont
  {Singer}(1973)}]{Herczeg:1974ik}%
  \BibitemOpen
  \bibfield  {author} {\bibinfo {author} {\bibfnamefont {P.}~\bibnamefont
  {Herczeg}}\ and\ \bibinfo {author} {\bibfnamefont {P.}~\bibnamefont
  {Singer}},\ }\href {\doibase 10.1103/PhysRevD.8.4107} {\bibfield  {journal}
  {\bibinfo  {journal} {Phys. Rev. D}\ }\textbf {\bibinfo {volume} {8}},\
  \bibinfo {pages} {4107} (\bibinfo {year} {1973})}\BibitemShut {NoStop}%
\bibitem [{\citenamefont {Geng}\ \emph {et~al.}(2002)\citenamefont {Geng},
  \citenamefont {NG},\ and\ \citenamefont {WU}}]{GENG_2002}%
  \BibitemOpen
  \bibfield  {author} {\bibinfo {author} {\bibfnamefont {C.~Q.}\ \bibnamefont
  {Geng}}, \bibinfo {author} {\bibfnamefont {J.~N.}\ \bibnamefont {NG}}, \ and\
  \bibinfo {author} {\bibfnamefont {T.~H.}\ \bibnamefont {WU}},\ }\href
  {\doibase 10.1142/s0217732302007697} {\bibfield  {journal} {\bibinfo
  {journal} {Modern Physics Letters A}\ }\textbf {\bibinfo {volume} {17}},\
  \bibinfo {pages} {1489–1497} (\bibinfo {year} {2002})}\BibitemShut
  {NoStop}%
\bibitem [{\citenamefont {Gao}(2002)}]{GAO_2002}%
  \BibitemOpen
  \bibfield  {author} {\bibinfo {author} {\bibfnamefont {D.-N.}\ \bibnamefont
  {Gao}},\ }\href {\doibase 10.1142/s0217732302007739} {\bibfield  {journal}
  {\bibinfo  {journal} {Modern Physics Letters A}\ }\textbf {\bibinfo {volume}
  {17}},\ \bibinfo {pages} {1583–1588} (\bibinfo {year} {2002})}\BibitemShut
  {NoStop}%
\bibitem [{\citenamefont {Scherer}\ and\ \citenamefont
  {Schindler}(2012)}]{Scherer:2012xha}%
  \BibitemOpen
  \bibfield  {author} {\bibinfo {author} {\bibfnamefont {S.}~\bibnamefont
  {Scherer}}\ and\ \bibinfo {author} {\bibfnamefont {M.~R.}\ \bibnamefont
  {Schindler}},\ }\href {\doibase 10.1007/978-3-642-19254-8} {\emph {\bibinfo
  {title} {{A Primer for Chiral Perturbation Theory}}}},\ Vol.\ \bibinfo
  {volume} {830}\ (\bibinfo {year} {2012})\BibitemShut {NoStop}%
\bibitem [{\citenamefont {Goldstone}\ \emph {et~al.}(1962)\citenamefont
  {Goldstone}, \citenamefont {Salam},\ and\ \citenamefont
  {Weinberg}}]{Goldstone:1962}%
  \BibitemOpen
  \bibfield  {author} {\bibinfo {author} {\bibfnamefont {J.}~\bibnamefont
  {Goldstone}}, \bibinfo {author} {\bibfnamefont {A.}~\bibnamefont {Salam}}, \
  and\ \bibinfo {author} {\bibfnamefont {S.}~\bibnamefont {Weinberg}},\ }\href
  {\doibase 10.1103/PhysRev.127.965} {\bibfield  {journal} {\bibinfo  {journal}
  {Phys. Rev.}\ }\textbf {\bibinfo {volume} {127}},\ \bibinfo {pages} {965}
  (\bibinfo {year} {1962})}\BibitemShut {NoStop}%
\bibitem [{\citenamefont {'t~Hooft}(1976)}]{tHooft:1976}%
  \BibitemOpen
  \bibfield  {author} {\bibinfo {author} {\bibfnamefont {G.}~\bibnamefont
  {'t~Hooft}},\ }\href {\doibase 10.1103/PhysRevLett.37.8} {\bibfield
  {journal} {\bibinfo  {journal} {Phys. Rev. Lett.}\ }\textbf {\bibinfo
  {volume} {37}},\ \bibinfo {pages} {8} (\bibinfo {year} {1976})}\BibitemShut
  {NoStop}%
\bibitem [{\citenamefont {Witten}(1979{\natexlab{a}})}]{WITTEN1979269}%
  \BibitemOpen
  \bibfield  {author} {\bibinfo {author} {\bibfnamefont {E.}~\bibnamefont
  {Witten}},\ }\href {\doibase https://doi.org/10.1016/0550-3213(79)90031-2}
  {\bibfield  {journal} {\bibinfo  {journal} {Nuclear Physics B}\ }\textbf
  {\bibinfo {volume} {156}},\ \bibinfo {pages} {269} (\bibinfo {year}
  {1979}{\natexlab{a}})}\BibitemShut {NoStop}%
\bibitem [{\citenamefont {Veneziano}(1979)}]{VENEZIANO1979213}%
  \BibitemOpen
  \bibfield  {author} {\bibinfo {author} {\bibfnamefont {G.}~\bibnamefont
  {Veneziano}},\ }\href {\doibase https://doi.org/10.1016/0550-3213(79)90332-8}
  {\bibfield  {journal} {\bibinfo  {journal} {Nuclear Physics B}\ }\textbf
  {\bibinfo {volume} {159}},\ \bibinfo {pages} {213} (\bibinfo {year}
  {1979})}\BibitemShut {NoStop}%
\bibitem [{\citenamefont {Hooft}(1974)}]{HOOFT1974461}%
  \BibitemOpen
  \bibfield  {author} {\bibinfo {author} {\bibfnamefont {G.}~\bibnamefont
  {Hooft}},\ }\href {\doibase https://doi.org/10.1016/0550-3213(74)90154-0}
  {\bibfield  {journal} {\bibinfo  {journal} {Nuclear Physics B}\ }\textbf
  {\bibinfo {volume} {72}},\ \bibinfo {pages} {461} (\bibinfo {year}
  {1974})}\BibitemShut {NoStop}%
\bibitem [{\citenamefont {Witten}(1979{\natexlab{b}})}]{WITTEN197957}%
  \BibitemOpen
  \bibfield  {author} {\bibinfo {author} {\bibfnamefont {E.}~\bibnamefont
  {Witten}},\ }\href {\doibase https://doi.org/10.1016/0550-3213(79)90232-3}
  {\bibfield  {journal} {\bibinfo  {journal} {Nuclear Physics B}\ }\textbf
  {\bibinfo {volume} {160}},\ \bibinfo {pages} {57} (\bibinfo {year}
  {1979}{\natexlab{b}})}\BibitemShut {NoStop}%
\bibitem [{\citenamefont {Coleman}\ and\ \citenamefont
  {Witten}(1980)}]{Coleman:1980}%
  \BibitemOpen
  \bibfield  {author} {\bibinfo {author} {\bibfnamefont {S.}~\bibnamefont
  {Coleman}}\ and\ \bibinfo {author} {\bibfnamefont {E.}~\bibnamefont
  {Witten}},\ }\href {\doibase 10.1103/PhysRevLett.45.100} {\bibfield
  {journal} {\bibinfo  {journal} {Phys. Rev. Lett.}\ }\textbf {\bibinfo
  {volume} {45}},\ \bibinfo {pages} {100} (\bibinfo {year} {1980})}\BibitemShut
  {NoStop}%
\bibitem [{\citenamefont {Leutwyler}(1996{\natexlab{a}})}]{LEUTWYLER1996163}%
  \BibitemOpen
  \bibfield  {author} {\bibinfo {author} {\bibfnamefont {H.}~\bibnamefont
  {Leutwyler}},\ }\href {\doibase https://doi.org/10.1016/0370-2693(96)85876-X}
  {\bibfield  {journal} {\bibinfo  {journal} {Physics Letters B}\ }\textbf
  {\bibinfo {volume} {374}},\ \bibinfo {pages} {163} (\bibinfo {year}
  {1996}{\natexlab{a}})}\BibitemShut {NoStop}%
\bibitem [{\citenamefont {Herrera-Sikl\'{o}dy}\ \emph
  {et~al.}(1997)\citenamefont {Herrera-Sikl\'{o}dy}, \citenamefont {Latorre},
  \citenamefont {Pascual},\ and\ \citenamefont
  {Taron}}]{HERRERASIKLODY1997345}%
  \BibitemOpen
  \bibfield  {author} {\bibinfo {author} {\bibfnamefont {P.}~\bibnamefont
  {Herrera-Sikl\'{o}dy}}, \bibinfo {author} {\bibfnamefont {J.}~\bibnamefont
  {Latorre}}, \bibinfo {author} {\bibfnamefont {P.}~\bibnamefont {Pascual}}, \
  and\ \bibinfo {author} {\bibfnamefont {J.}~\bibnamefont {Taron}},\ }\href
  {\doibase https://doi.org/10.1016/S0550-3213(97)00260-5} {\bibfield
  {journal} {\bibinfo  {journal} {Nuclear Physics B}\ }\textbf {\bibinfo
  {volume} {497}},\ \bibinfo {pages} {345} (\bibinfo {year}
  {1997})}\BibitemShut {NoStop}%
\bibitem [{\citenamefont {Kaiser}\ and\ \citenamefont
  {Leutwyler}(2000)}]{Kaiser:2000gs}%
  \BibitemOpen
  \bibfield  {author} {\bibinfo {author} {\bibfnamefont {R.}~\bibnamefont
  {Kaiser}}\ and\ \bibinfo {author} {\bibfnamefont {H.}~\bibnamefont
  {Leutwyler}},\ }\href@noop {} {\bibfield  {journal} {\bibinfo  {journal}
  {Eur. Phys. J. C}\ }\textbf {\bibinfo {volume} {17}},\ \bibinfo {pages} {623}
  (\bibinfo {year} {2000})}\BibitemShut {NoStop}%
\bibitem [{\citenamefont {Gasser}\ and\ \citenamefont
  {Leutwyler}(1985)}]{GASSER1985465}%
  \BibitemOpen
  \bibfield  {author} {\bibinfo {author} {\bibfnamefont {J.}~\bibnamefont
  {Gasser}}\ and\ \bibinfo {author} {\bibfnamefont {H.}~\bibnamefont
  {Leutwyler}},\ }\href {\doibase https://doi.org/10.1016/0550-3213(85)90492-4}
  {\bibfield  {journal} {\bibinfo  {journal} {Nuclear Physics B}\ }\textbf
  {\bibinfo {volume} {250}},\ \bibinfo {pages} {465} (\bibinfo {year}
  {1985})}\BibitemShut {NoStop}%
\bibitem [{\citenamefont {Leutwyler}(1996{\natexlab{b}})}]{Leutwyler:1996sa}%
  \BibitemOpen
  \bibfield  {author} {\bibinfo {author} {\bibfnamefont {H.}~\bibnamefont
  {Leutwyler}},\ }\href {\doibase https://doi.org/10.1016/0370-2693(96)85876-X}
  {\bibfield  {journal} {\bibinfo  {journal} {Physics Letters B}\ }\textbf
  {\bibinfo {volume} {374}},\ \bibinfo {pages} {163} (\bibinfo {year}
  {1996}{\natexlab{b}})}\BibitemShut {NoStop}%
\bibitem [{\citenamefont {Bickert}\ and\ \citenamefont
  {Scherer}(2020)}]{bickert2020}%
  \BibitemOpen
  \bibfield  {author} {\bibinfo {author} {\bibfnamefont {P.}~\bibnamefont
  {Bickert}}\ and\ \bibinfo {author} {\bibfnamefont {S.}~\bibnamefont
  {Scherer}},\ }\href {\doibase 10.1103/PhysRevD.102.074019} {\bibfield
  {journal} {\bibinfo  {journal} {Phys. Rev. D}\ }\textbf {\bibinfo {volume}
  {102}},\ \bibinfo {pages} {074019} (\bibinfo {year} {2020})}\BibitemShut
  {NoStop}%
\bibitem [{\citenamefont {B\"ar}\ and\ \citenamefont
  {Wiese}(2001)}]{Bar:2001qk}%
  \BibitemOpen
  \bibfield  {author} {\bibinfo {author} {\bibfnamefont {O.}~\bibnamefont
  {B\"ar}}\ and\ \bibinfo {author} {\bibfnamefont {U.-J.}\ \bibnamefont
  {Wiese}},\ }\href {\doibase https://doi.org/10.1016/S0550-3213(01)00288-7}
  {\bibfield  {journal} {\bibinfo  {journal} {Nuclear Physics B}\ }\textbf
  {\bibinfo {volume} {609}},\ \bibinfo {pages} {225} (\bibinfo {year}
  {2001})}\BibitemShut {NoStop}%
\bibitem [{\citenamefont {G\'{e}rard}\ and\ \citenamefont
  {Lahna}(1995)}]{Gerard:1995bv}%
  \BibitemOpen
  \bibfield  {author} {\bibinfo {author} {\bibfnamefont {J.-M.}\ \bibnamefont
  {G\'{e}rard}}\ and\ \bibinfo {author} {\bibfnamefont {T.}~\bibnamefont
  {Lahna}},\ }\href {\doibase https://doi.org/10.1016/0370-2693(95)00811-X}
  {\bibfield  {journal} {\bibinfo  {journal} {Physics Letters B}\ }\textbf
  {\bibinfo {volume} {356}},\ \bibinfo {pages} {381} (\bibinfo {year}
  {1995})}\BibitemShut {NoStop}%
\bibitem [{\citenamefont {Mertig}\ \emph {et~al.}(1991)\citenamefont {Mertig},
  \citenamefont {B\"ohm},\ and\ \citenamefont {Denner}}]{Mertig:1990an}%
  \BibitemOpen
  \bibfield  {author} {\bibinfo {author} {\bibfnamefont {R.}~\bibnamefont
  {Mertig}}, \bibinfo {author} {\bibfnamefont {M.}~\bibnamefont {B\"ohm}}, \
  and\ \bibinfo {author} {\bibfnamefont {A.}~\bibnamefont {Denner}},\ }\href
  {\doibase https://doi.org/10.1016/0010-4655(91)90130-D} {\bibfield  {journal}
  {\bibinfo  {journal} {Computer Physics Communications}\ }\textbf {\bibinfo
  {volume} {64}},\ \bibinfo {pages} {345} (\bibinfo {year} {1991})}\BibitemShut
  {NoStop}%
\bibitem [{\citenamefont {Bickert}\ \emph {et~al.}(2017)\citenamefont
  {Bickert}, \citenamefont {Masjuan},\ and\ \citenamefont
  {Scherer}}]{Bickert:2016fgy}%
  \BibitemOpen
  \bibfield  {author} {\bibinfo {author} {\bibfnamefont {P.}~\bibnamefont
  {Bickert}}, \bibinfo {author} {\bibfnamefont {P.}~\bibnamefont {Masjuan}}, \
  and\ \bibinfo {author} {\bibfnamefont {S.}~\bibnamefont {Scherer}},\ }\href
  {\doibase 10.1103/PhysRevD.95.054023} {\bibfield  {journal} {\bibinfo
  {journal} {Phys. Rev. D}\ }\textbf {\bibinfo {volume} {95}},\ \bibinfo
  {pages} {054023} (\bibinfo {year} {2017})}\BibitemShut {NoStop}%
\bibitem [{\citenamefont {Hacker}(2008)}]{Hacker:2008}%
  \BibitemOpen
  \bibfield  {author} {\bibinfo {author} {\bibfnamefont {C.}~\bibnamefont
  {Hacker}},\ }\href@noop {} {Ph.D. thesis},\ \bibinfo  {school} {Johannes
  Gutenberg-Universit\"at, Mainz} (\bibinfo {year} {2008}),\ \bibinfo {note}
  {http://nbn-resolving.org/urn:nbn:de:hebis:77-18401}\BibitemShut {NoStop}%
\bibitem [{\citenamefont {Zyla}\ \emph {et~al.}(2020)\citenamefont {Zyla} \emph
  {et~al.}}]{Zyla:2020zbs}%
  \BibitemOpen
  \bibfield  {author} {\bibinfo {author} {\bibfnamefont {P.}~\bibnamefont
  {Zyla}} \emph {et~al.} (\bibinfo {collaboration} {Particle Data Group}),\
  }\href {\doibase 10.1093/ptep/ptaa104} {\bibfield  {journal} {\bibinfo
  {journal} {PTEP}\ }\textbf {\bibinfo {volume} {2020}},\ \bibinfo {pages}
  {083C01} (\bibinfo {year} {2020})}\BibitemShut {NoStop}%
\bibitem [{\citenamefont {Adlarson}\ \emph {et~al.}(2012)\citenamefont
  {Adlarson} \emph {et~al.}}]{Adlarson:2011xb}%
  \BibitemOpen
  \bibfield  {author} {\bibinfo {author} {\bibfnamefont {P.}~\bibnamefont
  {Adlarson}} \emph {et~al.},\ }\href {\doibase
  https://doi.org/10.1016/j.physletb.2011.12.027} {\bibfield  {journal}
  {\bibinfo  {journal} {Physics Letters B}\ }\textbf {\bibinfo {volume}
  {707}},\ \bibinfo {pages} {243} (\bibinfo {year} {2012})}\BibitemShut
  {NoStop}%
\bibitem [{\citenamefont {Ablikim}\ \emph {et~al.}(2018)\citenamefont {Ablikim}
  \emph {et~al.}}]{Ablikim:2017fll}%
  \BibitemOpen
  \bibfield  {author} {\bibinfo {author} {\bibfnamefont {M.}~\bibnamefont
  {Ablikim}} \emph {et~al.} (\bibinfo {collaboration} {BESIII}),\ }\href
  {\doibase 10.1103/PhysRevLett.120.242003} {\bibfield  {journal} {\bibinfo
  {journal} {Phys. Rev. Lett.}\ }\textbf {\bibinfo {volume} {120}},\ \bibinfo
  {pages} {242003} (\bibinfo {year} {2018})}\BibitemShut {NoStop}%
\bibitem [{\citenamefont {Bijnens}\ \emph {et~al.}(1990)\citenamefont
  {Bijnens}, \citenamefont {Bramon},\ and\ \citenamefont
  {Cornet}}]{Bijnens:1989ff}%
  \BibitemOpen
  \bibfield  {author} {\bibinfo {author} {\bibfnamefont {J.}~\bibnamefont
  {Bijnens}}, \bibinfo {author} {\bibfnamefont {A.}~\bibnamefont {Bramon}}, \
  and\ \bibinfo {author} {\bibfnamefont {F.}~\bibnamefont {Cornet}},\ }\href
  {\doibase 10.1016/0370-2693(90)91212-T} {\bibfield  {journal} {\bibinfo
  {journal} {Phys. Lett. B}\ }\textbf {\bibinfo {volume} {237}},\ \bibinfo
  {pages} {488} (\bibinfo {year} {1990})}\BibitemShut {NoStop}%
\bibitem [{\citenamefont {Borasoy}\ and\ \citenamefont
  {Ni\ss{}ler}(2004)}]{Borasoy:2004qj}%
  \BibitemOpen
  \bibfield  {author} {\bibinfo {author} {\bibfnamefont {B.}~\bibnamefont
  {Borasoy}}\ and\ \bibinfo {author} {\bibfnamefont {R.}~\bibnamefont
  {Ni\ss{}ler}},\ }\href {http://dx.doi.org/10.1016/j.nuclphysa.2004.05.006}
  {\bibfield  {journal} {\bibinfo  {journal} {Nuclear Physics A}\ }\textbf
  {\bibinfo {volume} {740}},\ \bibinfo {pages} {362–382} (\bibinfo {year}
  {2004})}\BibitemShut {NoStop}%
\bibitem [{\citenamefont {Picciotto}(1992)}]{Picciotto:1991ae}%
  \BibitemOpen
  \bibfield  {author} {\bibinfo {author} {\bibfnamefont {C.}~\bibnamefont
  {Picciotto}},\ }\href {\doibase 10.1103/PhysRevD.45.1569} {\bibfield
  {journal} {\bibinfo  {journal} {Phys. Rev. D}\ }\textbf {\bibinfo {volume}
  {45}},\ \bibinfo {pages} {1569} (\bibinfo {year} {1992})}\BibitemShut
  {NoStop}%
\bibitem [{\citenamefont {Benayoun}\ \emph {et~al.}(2003)\citenamefont
  {Benayoun}, \citenamefont {David}, \citenamefont {DelBuono}, \citenamefont
  {Leruste},\ and\ \citenamefont {O\'{}Connell}}]{Benayoun:2003we}%
  \BibitemOpen
  \bibfield  {author} {\bibinfo {author} {\bibfnamefont {M.}~\bibnamefont
  {Benayoun}}, \bibinfo {author} {\bibfnamefont {P.}~\bibnamefont {David}},
  \bibinfo {author} {\bibfnamefont {L.}~\bibnamefont {DelBuono}}, \bibinfo
  {author} {\bibfnamefont {P.}~\bibnamefont {Leruste}}, \ and\ \bibinfo
  {author} {\bibfnamefont {H.~B.}\ \bibnamefont {O\'{}Connell}},\ }\href
  {https://doi.org/10.1140/epjc/s2003-01378-x} {\bibfield  {journal} {\bibinfo
  {journal} {Eur. Phys. J. C}\ }\textbf {\bibinfo {volume} {31}},\ \bibinfo
  {pages} {525} (\bibinfo {year} {2003})}\BibitemShut {NoStop}%
\bibitem [{\citenamefont {Benayoun}\ \emph {et~al.}(2010)\citenamefont
  {Benayoun}, \citenamefont {David}, \citenamefont {DelBuono},\ and\
  \citenamefont {Leitner}}]{Benayoun:2009im}%
  \BibitemOpen
  \bibfield  {author} {\bibinfo {author} {\bibfnamefont {M.}~\bibnamefont
  {Benayoun}}, \bibinfo {author} {\bibfnamefont {P.}~\bibnamefont {David}},
  \bibinfo {author} {\bibfnamefont {L.}~\bibnamefont {DelBuono}}, \ and\
  \bibinfo {author} {\bibfnamefont {O.}~\bibnamefont {Leitner}},\ }\href
  {https://doi.org/10.1140/epjc/s10052-009-1197-6} {\bibfield  {journal}
  {\bibinfo  {journal} {Eur. Phys. J. C}\ }\textbf {\bibinfo {volume} {65}},\
  \bibinfo {pages} {211} (\bibinfo {year} {2010})}\BibitemShut {NoStop}%
\bibitem [{\citenamefont {Osipov}\ \emph {et~al.}(2020)\citenamefont {Osipov},
  \citenamefont {Pivovarov}, \citenamefont {Volkov},\ and\ \citenamefont
  {Khalifa}}]{Osipov:2020vad}%
  \BibitemOpen
  \bibfield  {author} {\bibinfo {author} {\bibfnamefont {A.~A.}\ \bibnamefont
  {Osipov}}, \bibinfo {author} {\bibfnamefont {A.~A.}\ \bibnamefont
  {Pivovarov}}, \bibinfo {author} {\bibfnamefont {M.~K.}\ \bibnamefont
  {Volkov}}, \ and\ \bibinfo {author} {\bibfnamefont {M.~M.}\ \bibnamefont
  {Khalifa}},\ }\href {\doibase 10.1103/PhysRevD.101.094031} {\bibfield
  {journal} {\bibinfo  {journal} {Phys. Rev. D}\ }\textbf {\bibinfo {volume}
  {101}},\ \bibinfo {pages} {094031} (\bibinfo {year} {2020})}\BibitemShut
  {NoStop}%
\bibitem [{\citenamefont {Venugopal}\ and\ \citenamefont
  {Holstein}(1998)}]{Venugopal:1998fq}%
  \BibitemOpen
  \bibfield  {author} {\bibinfo {author} {\bibfnamefont {E.~P.}\ \bibnamefont
  {Venugopal}}\ and\ \bibinfo {author} {\bibfnamefont {B.~R.}\ \bibnamefont
  {Holstein}},\ }\href {\doibase 10.1103/PhysRevD.57.4397} {\bibfield
  {journal} {\bibinfo  {journal} {Phys. Rev. D}\ }\textbf {\bibinfo {volume}
  {57}},\ \bibinfo {pages} {4397} (\bibinfo {year} {1998})}\BibitemShut
  {NoStop}%
\bibitem [{\citenamefont {Holstein}(2002)}]{Holstein:2001bt}%
  \BibitemOpen
  \bibfield  {author} {\bibinfo {author} {\bibfnamefont {B.~R.}\ \bibnamefont
  {Holstein}},\ }\href {\doibase 10.1238/physica.topical.099a00055} {\bibfield
  {journal} {\bibinfo  {journal} {Physica Scripta}\ }\textbf {\bibinfo {volume}
  {T99}},\ \bibinfo {pages} {55} (\bibinfo {year} {2002})}\BibitemShut
  {NoStop}%
\bibitem [{\citenamefont {Dai}\ \emph {et~al.}(2018)\citenamefont {Dai},
  \citenamefont {Kang}, \citenamefont {Mei\ss{}ner}, \citenamefont {Song},\
  and\ \citenamefont {Yao}}]{Dai:2017tew}%
  \BibitemOpen
  \bibfield  {author} {\bibinfo {author} {\bibfnamefont {L.-Y.}\ \bibnamefont
  {Dai}}, \bibinfo {author} {\bibfnamefont {X.-W.}\ \bibnamefont {Kang}},
  \bibinfo {author} {\bibfnamefont {U.-G.}\ \bibnamefont {Mei\ss{}ner}},
  \bibinfo {author} {\bibfnamefont {X.-Y.}\ \bibnamefont {Song}}, \ and\
  \bibinfo {author} {\bibfnamefont {D.-L.}\ \bibnamefont {Yao}},\ }\href
  {\doibase 10.1103/PhysRevD.97.036012} {\bibfield  {journal} {\bibinfo
  {journal} {Phys. Rev. D}\ }\textbf {\bibinfo {volume} {97}},\ \bibinfo
  {pages} {036012} (\bibinfo {year} {2018})}\BibitemShut {NoStop}%
\bibitem [{\citenamefont {Picciotto}\ and\ \citenamefont
  {Richardson}(1993)}]{Picciotto:1993aa}%
  \BibitemOpen
  \bibfield  {author} {\bibinfo {author} {\bibfnamefont {C.}~\bibnamefont
  {Picciotto}}\ and\ \bibinfo {author} {\bibfnamefont {S.}~\bibnamefont
  {Richardson}},\ }\href {\doibase 10.1103/PhysRevD.48.3395} {\bibfield
  {journal} {\bibinfo  {journal} {Phys. Rev. D}\ }\textbf {\bibinfo {volume}
  {48}},\ \bibinfo {pages} {3395} (\bibinfo {year} {1993})}\BibitemShut
  {NoStop}%
\bibitem [{\citenamefont {Borasoy}\ and\ \citenamefont
  {Ni\ss{}ler}(2007)}]{Borasoy:2007dw}%
  \BibitemOpen
  \bibfield  {author} {\bibinfo {author} {\bibfnamefont {B.}~\bibnamefont
  {Borasoy}}\ and\ \bibinfo {author} {\bibfnamefont {R.}~\bibnamefont
  {Ni\ss{}ler}},\ }\href {http://dx.doi.org/10.1140/epja/i2007-10396-3}
  {\bibfield  {journal} {\bibinfo  {journal} {Eur.~Phys.~J.~A}\ }\textbf
  {\bibinfo {volume} {33}},\ \bibinfo {pages} {95} (\bibinfo {year}
  {2007})}\BibitemShut {NoStop}%
\bibitem [{\citenamefont {Ablikim}\ \emph {et~al.}(2021)\citenamefont {Ablikim}
  \emph {et~al.}}]{Ablikim:2020svz}%
  \BibitemOpen
  \bibfield  {author} {\bibinfo {author} {\bibfnamefont {M.}~\bibnamefont
  {Ablikim}} \emph {et~al.} (\bibinfo {collaboration} {BESIII}),\ }\href
  {\doibase 10.1103/PhysRevD.103.072006} {\bibfield  {journal} {\bibinfo
  {journal} {Phys. Rev. D}\ }\textbf {\bibinfo {volume} {103}},\ \bibinfo
  {pages} {072006} (\bibinfo {year} {2021})}\BibitemShut {NoStop}%
\bibitem [{\citenamefont {Faessler}\ \emph {et~al.}(2000)\citenamefont
  {Faessler}, \citenamefont {Fuchs},\ and\ \citenamefont
  {Krivoruchenko}}]{Faessler:1999de}%
  \BibitemOpen
  \bibfield  {author} {\bibinfo {author} {\bibfnamefont {A.}~\bibnamefont
  {Faessler}}, \bibinfo {author} {\bibfnamefont {C.}~\bibnamefont {Fuchs}}, \
  and\ \bibinfo {author} {\bibfnamefont {M.~I.}\ \bibnamefont
  {Krivoruchenko}},\ }\href {http://dx.doi.org/10.1103/PhysRevC.61.035206}
  {\bibfield  {journal} {\bibinfo  {journal} {Phys.~Rev.~C}\ }\textbf {\bibinfo
  {volume} {61}},\ \bibinfo {pages} {035206} (\bibinfo {year}
  {2000})}\BibitemShut {NoStop}%
\bibitem [{\citenamefont {Petri}(2010)}]{Petri:2010ea}%
  \BibitemOpen
  \bibfield  {author} {\bibinfo {author} {\bibfnamefont {T.}~\bibnamefont
  {Petri}},\ }\href@noop {} {\enquote {\bibinfo {title} {Anomalous decays of
  pseudoscalar mesons},}\ } (\bibinfo {year} {2010}),\ \Eprint
  {http://arxiv.org/abs/1010.2378} {arXiv:1010.2378 [nucl-th]} \BibitemShut
  {NoStop}%
\end{thebibliography}%

\end{document}